\documentclass[aps,pre,reprint,superscriptaddress,nofootinbib, twocolumn]{revtex4-1}
\usepackage{graphicx}
\usepackage{amsmath, amssymb, amscd, latexsym, bm, braket}
\usepackage{hyperref}
\hypersetup{colorlinks=true, citecolor=blue, filecolor= black, linkcolor= blue, urlcolor= blue}

\begin{document}

\title{Uniaxial stress controlled anisotropic Rashba effects and carriers-based currents in BiTeI monolayer semiconductor}

\author{Shi-Hao Zhang}
\affiliation{Beijing National Laboratory for Condensed Matter Physics, Institute of Physics, Chinese Academy of Sciences, Beijing 100190, China}
\affiliation{School of Physical Sciences, University of Chinese Academy of Sciences, Beijing 100190, China}
\author{Bang-Gui Liu}
\email[]{bgliu@iphy.ac.cn}
\affiliation{Beijing National Laboratory for Condensed Matter Physics, Institute of Physics, Chinese Academy of Sciences, Beijing 100190, China}
\affiliation{School of Physical Sciences, University of Chinese Academy of Sciences, Beijing 100190, China}

\date{\today}

\begin{abstract}
Manipulation of Rashba effects in two-dimensional (2D) electron systems is highly desirable for controllable applications in spintronics and optoelectronics. Here, by combining first-principles investigation and model analysis, we use uniaxial stress to control BiTeI monolayer as a Rashba 2D semiconductor for useful spin and transport properties.
We find that the stress-driven electron system can be described by an effective anisotropic Rashba model including all the three Pauli matrixes, and uniaxial stress allows an out-of-plane spin component.
When appropriate electron carriers are introduced into the monolayer, an in-plane electric field can induce a charge current and three spin current components (including that based on the out-of-plane spin) because of the reduced symmetry. Therefore, uniaxial stress can be used to control such Rashba 2D electron systems as the BiTeI monolayer for seeking promising devices.
\end{abstract}


\maketitle


\section{Introduction}

Since the advent of graphene, it becomes clear that two-dimensional (2D) materials can be used to realize high-performance devices for next-generation electronic and optical applications. Various important electronic, magnetic, optical, and mechanical phenomena and effects have been observed. As an important effect,
Rashba effect is a spin splitting phenomenon, originating from the spin-orbit coupling (SOC) in electronic systems without out-of-plane mirror symmetry~\cite{Rashba1959,Rashba1960,SpinSec2017}. In a 2D electron gas system, this phenomenon can be descried by the famous Rashba model~\cite{bychkov1984properties,Manchon2015}.
The energy bands of this two-band model have a crossing point at the $\Gamma$ point.
The extremum of the two bands is located along a circle of radius $k_0$ around the $\Gamma$ point, and the energy difference between the extremum and the crossing point is defined as Rashba splitting energy $\rm \mathcal{E}_R$~\cite{SpinSec2017}. It has been proved through first-principles or experimental investigations that there exist strong Rashba effects in some monolayers, multilayer, and heterostructures, such as Janus transition-metal dichalcogenide monolayers~\cite{PhysRevB.97.235404}, BiSb monolayers~\cite{PhysRevB.95.165444,YUAN2018163,Lu2017}, PbX monolayers (X=S, Se, Te)~\cite{PhysRevB.96.161401,PhysRevB.97.235312}, Ag$_2$Te monolayer~\cite{ag2te}, gated multilayer InSe~\cite{acs.nanolett.8b01462}, GaX/MoX$_2$ (X=S, Se, Te) heterostructures~\cite{PhysRevB.97.155415}.
The BiTeI bulk, as a polar crystal with layered crystal structure, has attracted more and more attention because of the strong Rashba splitting~\cite{Ishizaka2011,PhysRevLett.109.096803,PhysRevLett.109.116403,PhysRevLett.108.246802,PhysRevLett.110.107204,PhysRevMaterials.1.054201,PhysRevLett.111.166403,PhysRevB.86.085204,Butler2014,PhysRevB.91.245312} , optical response~\cite{PhysRevLett.107.117401,PhysRevLett.109.167401}, and topological physics~\cite{Bahramy2012,PhysRevB.96.155309,PhysRevLett.121.246403,10.1002/adma.201605965}. The BiTeI monolayer is also predicted as a polar material with giant Rashba effect~\cite{ma2014emergence}. Recently, the BiTeI monolayer was synthesised experimentally~\cite{fulop2018exfoliation}. It is highly desirable to manipulate the BiTeI monolayer by applying uniaxial stress in order to take full potential of the BiTeI monolayer for controlling the Rashba effects and seeking promising charge/spin currents.

Here, we use uniaxial stress to manipulate the BiTeI monolayer semiconductor through combining first-principles investigations and theoretical model analyses. Our first-principles results reveal that uniaxial stress can cause strong anisotropy in the energy bands near the $\Gamma$ point, and makes out-of-plane spin component occur. We obtain an effective two-band Hamiltonian to describe the stress-dependent conduction bands and spin texture well. After introducing electron carriers of low concentration, an in-plane electric field can induce a charge current and three nonzero spin current components, including an out-of-plane spin current component, in the monolayer because of the broken inversion symmetry. Controllable anisotropic Rashba effects and carriers-based charge/spin currents can be realized in this way. More detailed results will be presented in the following.

\section{Methodology}

The first-principles calculations are performed with the projector-augmented wave (PAW) method within the density functional theory~\cite{PhysRevB.50.17953}, as implemented in the Vienna ab initio simulation package software (VASP)~\cite{PhysRevB.47.558}. The kinetic energy cutoff of the plane waves is set to 400 eV. All atomic positions are fully optimized until the energy difference between two successive steps is smaller than 10$^{-6}$ eV and the Hellmann-Feynman forces on each atom are less than 0.01 eV/\AA{}. The generalized gradient approximation (GGA) by Perdew, Burke, and Ernzerhof (PBE)~\cite{PhysRevLett.77.3865} is used as the exchange-correlation functional. The Brillouin zone integration in the self-consistent calculation is carried out with a 42$\times$28$\times$1 special $\Gamma$-centered k-point mesh following the convention of Monkhorst-Pack~\cite{PhysRevB.13.5188}. The spin-orbit coupling (SOC) effect is taken into consideration in all calculations including structure optimizations, self-consistent calculations, and energy-bands calculations. In order to calculate Rashba coefficients, we take 1001 k-points along each high-symmetry line in the energy band calculations to avoid the errors coming from the discreteness of k-points and ensure the convergence of Rashba parameter along different directions.

\section{Result and discussion}

\subsection{Uniaxial stress and anisotropic Rashba effects}

\begin{figure}[!htbp]
\includegraphics[width=0.48\textwidth]{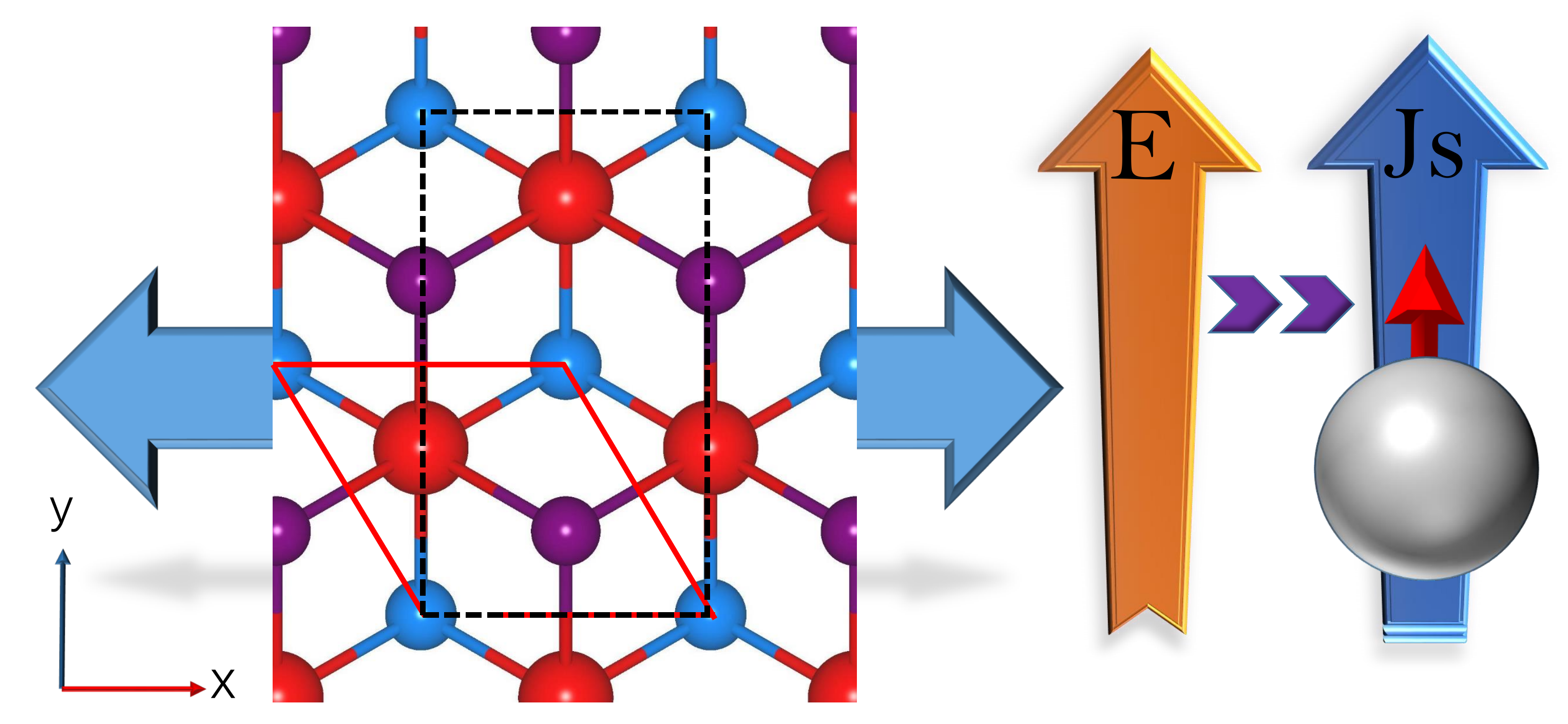}
\caption{~\label{fig1} The crystal structure of BiTeI monolayer and a illustration of the uniaxial stress. The red, blue, and purple balls refer to the bismuth, tellurium, and iodine atoms, respectively. When applying on the stretched monolayer, in-plane electric field can cause a spin current shown.}
\end{figure}

The structure of BiTeI monolayer, without external stress, is presented in Fig.~\ref{fig1}. The red solid line represents the unit cell of BiTeI monolayer. It obeys the P3m1 space group and shows the $C_{3v}$ symmetry. For convenience, we take the rectangular cell remarked by black dash line to calculate the effect of uniaxial stress. Through applying the uniaxial tensile stress upon the monolayer along x (or y) direction, the C$_{3v}$ symmetry will be removed and the monolayer keeps the $M_x$  mirror symmetry only, and there appears substantial structural anisotropy between the x and y directions. When a tensile uniaxial stress is along the x (y) axis, there will be a tensile strain along the x (y) axis and a compressive strain along the y (x) axis due to Poisson effect. Actually, we allow the tensile x (y) strain to change actively and determine the compressive y (x) strain by structural optimization, and then we estimate the tensile x (y) stress  by $\rm \partial \mathcal{E}/A\partial \eta$ where $\mathcal{E}$ is the total energy of the monolayer under $\rm \eta$ strain along the x (y) direction and $A$ is the area of the stretched unit cell. Our study show that the tensile 10\% strain along the x or y axis can be achieved by applying 1.71 N/m or 1.82 N/m as the uniaxial stress along the same axis, which indicates that the uniaxial stress on the BiTeI monolayer is experimentally accessible.

\begin{figure*}[!htbp]
\includegraphics[width=0.96\textwidth]{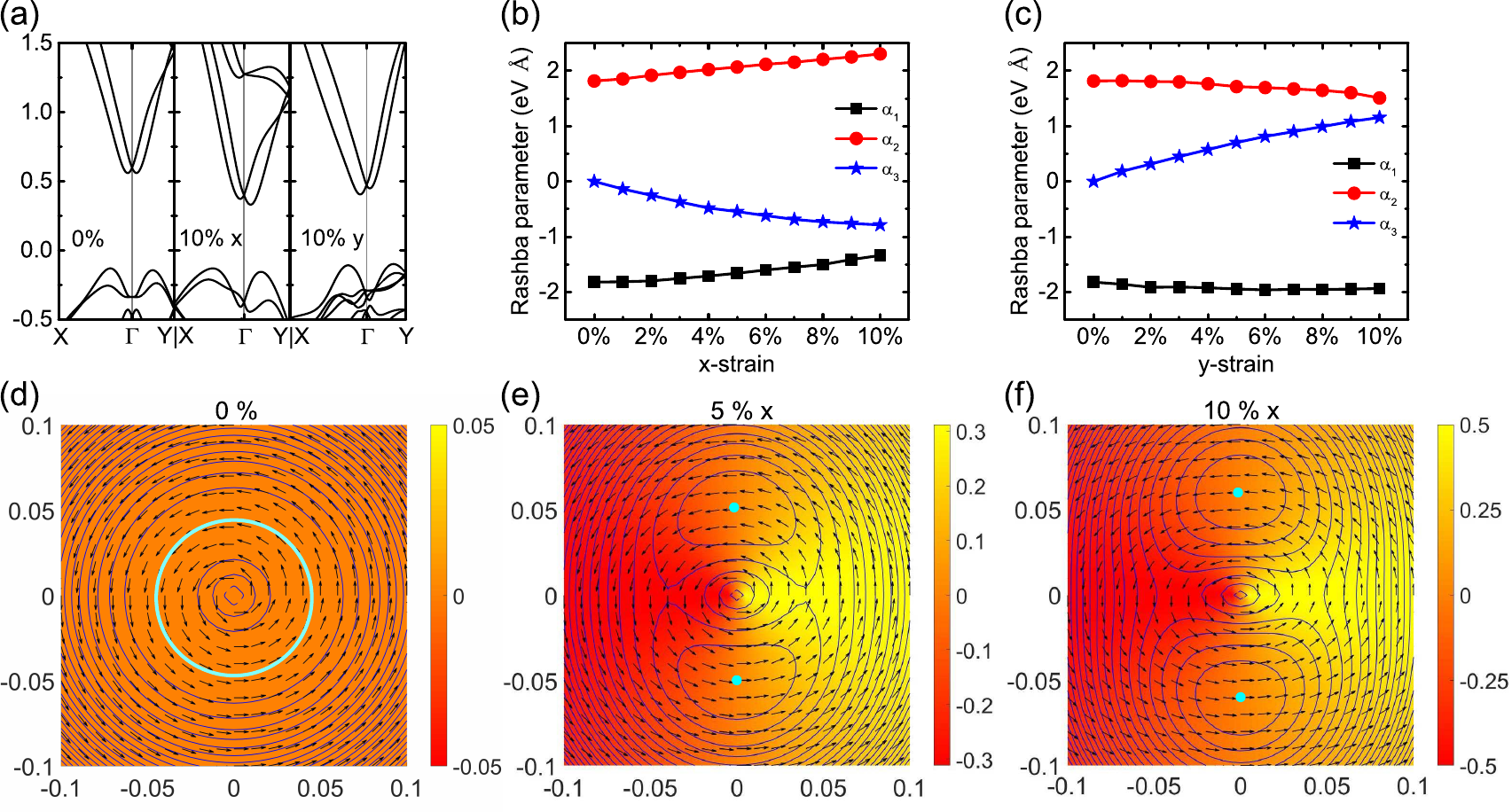}
\caption{~\label{fig2} (a) The energy bands of the monolayer with 0\%, 10\% x-strain, or 10\% y-strain. (b,c) The Rashba parameters ($\alpha_1$, $\alpha_2$, $\alpha_3$) in Hamiltonian (1) for different x- or y-strain values. (d-f) The spin textures of the lowest Rashba band near the $\Gamma$ point under 0\%, 5\%, or 10\% x-strains caused by uniaxial x-stress, where the in-plane spin components are indicated as arrows and the out-of-plane spin components are presented with the color scale. The contours of energy are shown by the blue lines and the energy minima are marked as the cyan circle or point.}
\end{figure*}

The effect of the uniaxial stress on energy bands of the BiTeI monolayer is shown in Fig. 2 (a), where the left part describes the bands without stress, the middle part those for 10\% tensile strain (stress 1.71 N/m) along the x axis, and the right part those for 10\% tensile strain (stress 1.82 N/m) along the y axis. When no stress is applied on the monolayer, the conduction bands near the $\Gamma$ point behave isotropic, and the calculated Rashba parameters ($\rm \mathcal{E}_R$, $k_0$, and $\alpha$) are 38.4 meV, 0.042 \AA{}$^{-1}$, and 1.82 eV \AA{}, in agreement with experimental values~\cite{fulop2018exfoliation}. When the tensile x-strain reaches 10\% (x-stress 1.71 N/m), the conduction band minima along the $k_y$ direction become much lower than those along the $k_x$ direction, which means that the conduction bands near the $\Gamma$ point become strongly anisotropic. As a result, the calculated Rashba splitting energy $\rm \mathcal{E}_R$ and k-vector off-set $k_0$ are equivalent to 27.5 meV and 0.035 \AA{}$^{-1}$ along the $k_x$ direction, but they become 69.6 meV and 0.060 \AA{}$^{-1}$ along the $k_y$ direction. When the tensile y-stress 1.82 N/m (10\% tensile strain) is applied, the evaluated parameters $\rm \mathcal{E}_R$ and $k_0$ are 23.6 meV and 0.031 \AA{}$^{-1}$ along the $k_y$ direction, but 67.7 meV and 0.060 \AA{}$^{-1}$ along the $k_x$ direction. These results reveal that the  uniaxial x (y) stress makes the lowest conduction bands around the $\Gamma$ point become two minimum points along the $\pm k_y$ ($\pm k_x$) direction, leading to strong anisotropy in the Rashba parameters.

When the uniaixial stress is applied, the rotational symmetry  no longer exists in the stretched BiTeI monolayer, and there remains only $\mathcal{M}_x$ mirror symmetry. Under the mirror transformation $\mathcal{M}_x$: $x \to -x$, $(k_x,k_y) \to (-k_x,k_y)$, and $(\sigma _x,\sigma _y,\sigma _z) \to (\sigma _x,-\sigma _y,-\sigma _z)$. Then we can construct the two-band Hamiltonian for the lowest conduction bands near the $\Gamma$ point,
\begin{eqnarray}
\hat{H} = \mathcal{E}_k+\alpha_1k_x\sigma_y+\alpha_2k_y\sigma_x+\alpha_3k_x\sigma_z,
\end{eqnarray}
where $\mathcal{E}_k=\frac{\hbar ^2}{2}(\frac{k^2_x}{m_x}+\frac{k^2_y}{m_y})$, and the crossing point of the two bands is located at $\mathcal{E}_k=0$. The three Rashba parameters $\alpha _1$, $\alpha _2$, and $\alpha _3$ in the Hamiltonian (1) can be obtained through fitting the conduction bands of strained monolayer, and the strain dependences of the parameters for the x/y stress are shown in Fig.~\ref{fig2} (b,c). It is interesting that $\alpha_3$  is negative for the uniaxial x-stress, but becomes positive for the uniaxial y stress. We can see that the uniaxial stress allows the emergence of new term $k_x\sigma_z$ in the Hamiltonian. It is a new physical phenomenon arising from structural symmetry $\mathcal{M}_x$. Considering the Rashba Hamiltonian with rotational group, there is no $\sigma _z$ term up to the third order of $k$ when the symmetry obeys the $C_2$, $C_{2v}$, $C_4$ and $C_{4v}$~\cite{PhysRevB.85.075404}. When the point symmetry is $C_3$ or $C_{3v}$, there is no $\sigma _z$ term in the first order of $k$, but some $\sigma _z$ terms in the third order of $k$ are allowed, for example, $(k_x^3-3k_xk_y^2)\sigma _z$ term~\cite{PhysRevB.85.075404,PhysRevLett.103.266801}. Thus we can conclude that uniaxial stress removes the $C_{3v}$ symmetry and thus leads to the emergence of the $k_x\sigma_z$ term, which can cause the inversion of out-of-plane spin polarization.

The eigenvalues of the two-band Hamiltonian (1) can be written as
\begin{eqnarray}
\mathcal{E}_k^{\lambda}=\mathcal{E}_k+\lambda \sqrt{(\alpha _1^2+\alpha _3^2)k_x^2+\alpha _2^2k_y^2},
\end{eqnarray}
where  $\lambda = \pm 1$ refers to the two bands. It should be noted that $\mathcal{E}_k=0$ means $\mathcal{E}^\lambda_k=0$. We can obtain the spin expectation values at the momentum $\mathbf{k}$ near the $\Gamma$ point,
\begin{eqnarray}
\langle \mathbf{k}\lambda|\sigma _x|\mathbf{k}\lambda \rangle=\frac{\lambda \alpha _2k_y}{\sqrt{(\alpha _1^2+\alpha _3^2)k_x^2+\alpha _2^2k_y^2}},\\
\langle \mathbf{k}\lambda|\sigma _y|\mathbf{k}\lambda \rangle=\frac{\lambda \alpha _1k_x}{\sqrt{(\alpha _1^2+\alpha _3^2)k_x^2+\alpha _2^2k_y^2}},\\
\langle \mathbf{k}\lambda|\sigma _z|\mathbf{k}\lambda \rangle=\frac{\lambda \alpha _3k_x}{\sqrt{(\alpha _1^2+\alpha _3^2)k_x^2+\alpha _2^2k_y^2}}.
\end{eqnarray}
The expression (5) means that there emerges a spin z component, in addition to the in-plane components. It is also indicated that the spin y-component and z-component along the $k_x = 0$ line are zero, which is in agreement with our effective Hamiltonian. This phenomenon can be regarded as the result of $\mathcal{M}_x$ mirror symmetry.
Our first-principles investigation about the spin texture of the lowest conduction band with 10\% strain along both x and y axes shows that the out-of-plane spin component has important effect.

Furthermore, we present both the spin textures and energy contours of the lower Rashba band ($\lambda=-1$) near the $\Gamma$ point for 0, 5\%, and 10\% x-strain in Fig.~\ref{fig2} (d-f). When the Fermi level is shifted downward from the crossing point, the Fermi lines will finally become two separated closed curves, even reduce to two points at the conduction band edge. The change of the Fermi line topology makes a Lifshitz transition. For low electron concentration, there are two Fermi pockets near the conduction band edge, as shown in Fig.~\ref{fig2} (e,f).

\begin{figure*}[!htbp]
\includegraphics[width=0.8\textwidth]{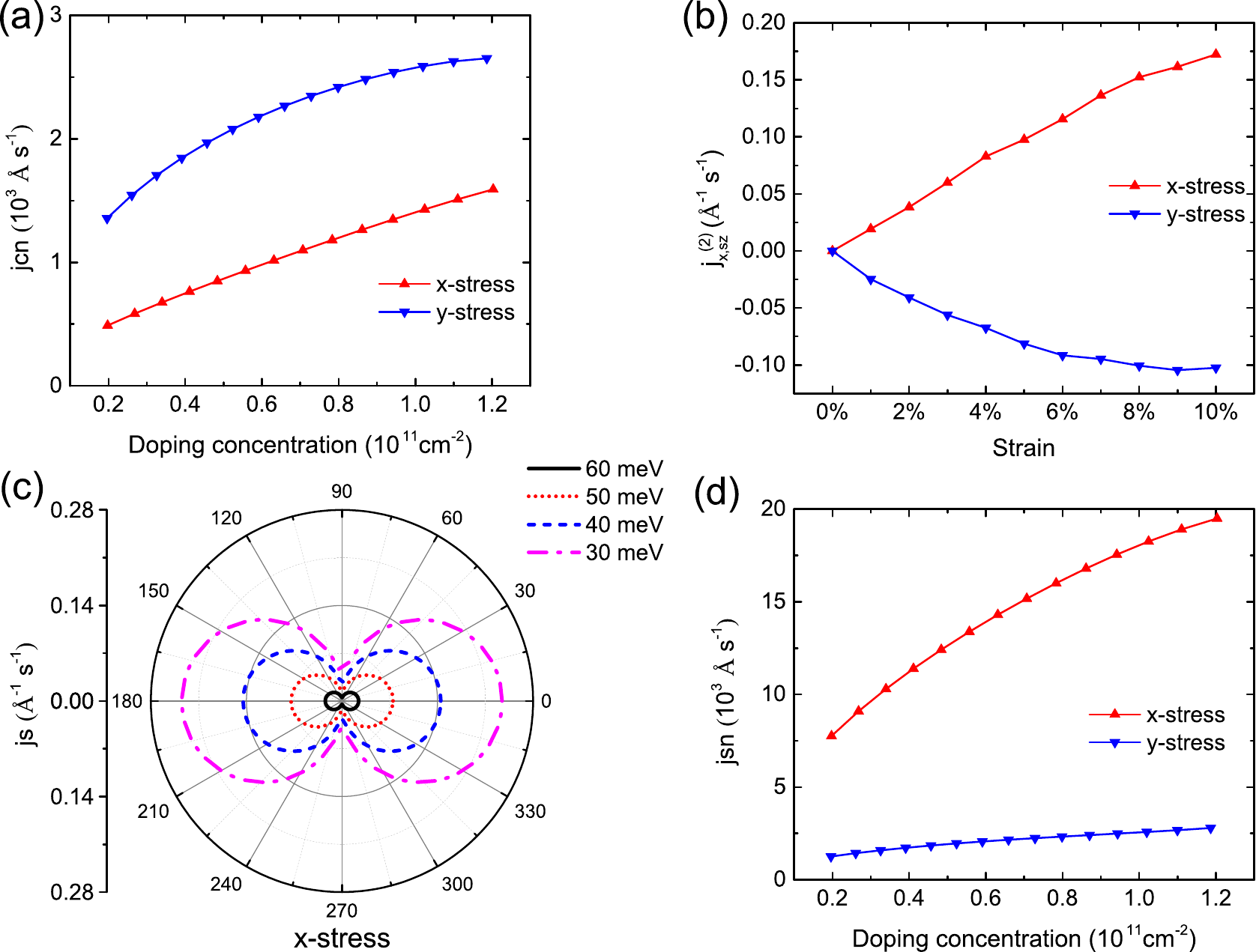}
\caption{~\label{fig3} (a) The normalized charge currents $j_{cn}$ in $e\tau E/\hbar ^2$ as functions of electron concentrations with 10\% x-strain or y-strain. (b) The second-order spin currents $j^{(2)}_{x, s_z}$ in $e^2\tau ^2E^2/\hbar ^3$ under different x-strain or y-strain caused by uniaxial stress, where the Fermi level is 10 meV lower than the crossing point. (c) The spin currents $j_s$ in $e^2\tau ^2E^2/\hbar ^3$ with 10\% x-strain, as functions of the $\theta$ angle, when the Fermi level is lower than the crossing point by 60 meV, 50 meV, 40 meV, or 30 meV. The conduction band energy minima are lower than the crossing point by 69.6 meV for 10\% x-strain, or 67.7 meV for 10\% y-strain. (d) The normalized maximal spin currents $j_{sn}$ in $e^2\tau ^2E^2/\hbar ^3$ as functions of electron concentrations under 10\% x-strain or y-strain. The electric field is along the x direction in all the cases.}
\end{figure*}

\subsection{Carrier charge and spin currents through electric field}

Now we address electric-field-induced transport phenomena in the stretched BiTeI monolayer in the presence of electron carriers, which can be achieved by charge transferring between van der Waals layers, electric gating, or chemical doping. We suppose that the in-plane electric field is so weak and the electron concentration is so small that there are little changes in the crystal structure and lower conduction bands of the BiTeI monolayer. By keeping in the regime of low electron concentration, we fix the Fermi level near the conduction band edge, as shown in Fig.~\ref{fig3}, and the occupied electrons are located in the lower Rashba band in the neighborhood of the $\Gamma$ points. When a small in-plane electric field $\mathbf{E}$ is applied, the relaxation time approximation for the Boltzmann equation of the distribution function $f$  can be expressed as
$-\frac{e\mathbf{E}}{\hbar}\cdot \frac{\partial f}{\partial \mathbf{k}}=-\frac{f-f_0}{\tau}$,
where $\tau$ is electron relaxation time and here Hall effect is not taken into consideration~\cite{PhysRevLett.113.156603,PhysRevB.95.224430}.
As the response to the external electric field, the distribution function can be expanded as $f=f_0+f_1+f_2+\cdots $, where $f_0$ is the distribution function of the system in the absence of $\mathbf{E}$, and $f_n = (\frac{e\tau}{\hbar} \mathbf{E}\cdot \frac{\partial }{\partial \mathbf{k}})^nf_0$ comes from the iterative substitution of the Boltzmann equation~\cite{PhysRevLett.115.216806,PhysRevB.94.245121,PhysRevLett.113.156603,PhysRevB.95.224430}.

Consequently, the charge current $\mathbf{j}_c$ can be written as~\cite{PhysRevLett.113.156603,PhysRevB.95.224430}
\begin{align}
\mathbf{j}_c&=-\frac{e\tau}{\hbar ^2}\int d\mathbf{k} \delta (\mathcal{E}^\lambda_k-\mathcal{E}_F)(\mathbf{E}\cdot \nabla _{k}\mathcal{E}^\lambda_k)\nabla _{k}\mathcal{E}^\lambda_k,
\end{align}
where $\lambda=-1$.
We can define a normalized charge current as the ratio of current magnitude to doping electron concentration to describe the current-production efficiency. When the electric field is applied along the x direction, the normalized charge current $j_{cn}$ as the function of electron concentrations under 10\% strain along x/y direction are shown in Fig.~\ref{fig3} (a). We can explain this phenomenon with the Drude model $j = ne^2\tau E/m^{*}$ where $n$ being the concentration of doped electrons. When Fermi level is shifted upward from the band edge, $\frac{\partial ^2 E}{\partial k_x^2}=\frac{\hbar ^2}{m_x}-(\alpha _1^2+\alpha _3^2)\alpha _2^2k_y^2/[(\alpha _1^2+\alpha _3^2)k_x^2+\alpha _2^2k_y^2]^{\frac{3}{2}}$ will be increasing, which reduces the $m^{*}$ value in the system. Thus the normalized current in $e\tau E$ will be monotonically increasing at this circumstance.

As for the spin current, we take the conventional definition of spin-current operator as
$\hat{J}_{\mu ,s_{\nu}}=\frac{1}{4} \{\frac{\partial \hat{H}}{\partial k_{\mu}},\sigma _{\nu}\}$,
which refers to the spin component $s_\nu$ flowing along the $\mu$ direction~\cite{PhysRevB.95.224430,PhysRevB.68.241315,PhysRevLett.92.126603}. It should be especially noted that this definition of spin-current operator will be invalid when the third-power k terms are included in the Hamiltonian~\cite{PhysRevB.83.113307}. The k terms in our Hamiltonian are only to the first order in the presence of small electron concentration ($\mathcal{E}_F < 0$).

Then we obtain the three non-zero components of the spin current near the $\Gamma$ point,
\begin{align}
\langle \mathbf{k}\lambda |\hat{J}_{y,s_x}|\mathbf{k}\lambda \rangle &=\frac{\lambda \hbar ^2\alpha _2k_y^2}{2m_y\sqrt{(\alpha _1^2+\alpha _3^2)k_x^2+\alpha _2^2k_y^2}}+\frac{\alpha _2}{2},\\
\langle \mathbf{k}\lambda |\hat{J}_{x,s_y}|\mathbf{k}\lambda \rangle &=\frac{\lambda \hbar ^2\alpha _1k_x^2}{2m_x\sqrt{(\alpha _1^2+\alpha _3^2)k_x^2+\alpha _2^2k_y^2}}+\frac{\alpha _1}{2},\\
\langle \mathbf{k}\lambda |\hat{J}_{x,s_z}|\mathbf{k}\lambda \rangle &=\frac{\lambda \hbar ^2\alpha _3k_x^2}{2m_x\sqrt{(\alpha _1^2+\alpha _3^2)k_x^2+\alpha _2^2k_y^2}}+\frac{\alpha _3}{2}.
\end{align}
For the isotropic Rashba model, we have $m_x = m_y$, $\alpha _1 = -\alpha _2$, and $\alpha _3=0$, and then derive $\langle \hat{J}_{y,s_x}\rangle = -\langle \hat{J}_{x,s_y}\rangle$, which is the same as Rashba's result~\cite{PhysRevB.68.241315}. The non-zero spin-current expectation will not lead to transport and accumulation of spin under the condition of thermodynamic equilibrium, and it can be connected to a non-Abelian SU(2) field generated by the spin-orbit coupling~\cite{PhysRevLett.101.106601}. It can have real spin transport  when the Rashba medium is constructed with a spatially modulated spin-orbit parameter~\cite{PhysRevB.76.033306,PhysRevB.77.035327}. For this reason, the zero-order spin current will be addressed no more in the following.

It has been shown that odd (even) orders of the spin (charge) current become zero in the presence of the time-reversal symmetry, and the broken inversion symmetry ensures the existence of non-zero even spin current~\cite{PhysRevB.95.224430}. So the second-order spin current in our system can occur responding to the electric field. The second-order spin current is defined by~\cite{PhysRevB.95.224430}
\begin{align}
j^{(2)}_{\mu ,s_{\nu}} & = \int \frac{d^2 \mathbf{k}}{(2\pi)^2} \langle \mathbf{k}\lambda |\hat{J}_{\mu ,s_{\nu}}|\mathbf{k}\lambda \rangle f_2 \\
 & = \frac{e^2\tau ^2}{4\pi ^2 \hbar ^3}\int d \mathbf{k} \delta (\mathcal{E}^\lambda_{k}-\mathcal{E}_F)(\mathbf{E}\cdot \nabla _{k}\mathcal{E}^\lambda_{k})\mathbf{E}\cdot \nabla _{\mathbf{k}}\langle \mathbf{k}\lambda |\hat{J}_{\mu ,s_{\nu}}|\mathbf{k}\lambda \rangle
\end{align}
When the electric field $\mathbf{E}$ is applied along the x or y direction, there are only three non-zero components for the spin currents, $j^{(2)}_{x, s_y}$, $j^{(2)}_{x, s_z}$ and $j^{(2)}_{y, s_x}$. Returning to the isotropic Rashba system, our numerical calculations give that $j^{(2)}_{x, s_y}=5j^{(2)}_{y, s_x}$ and $j^{(2)}_{x, s_z}=0$, which is in agreement with spin current results of isotropic Rashba model~\cite{PhysRevB.95.224430}. In Fig.~\ref{fig3} (b), we present the spin current $j^{(2)}_{x, s_z}$ under uniaxial stress (0$\sim$10\%) along x/y direction when the Fermi level is 10 meV lower than crossing point. When the x-strain (y-strain) reached 10\%, the magnitude of $j^{(2)}_{x, s_z}$ is equivalent to $|j^{(2)}_{x, s_z}| \approx 0.6|j^{(2)}_{x, s_y}|$. It is clear that the uniaxial stress leads to remarkable spin $s_z$ current $j^{(2)}_{x, s_z}$, in contrast to the isotropic case.

It is necessary to clarify the real-space texture of the electrically-generated spin current. The magnitude of the spin current along $\theta$ direction ($\theta=0$ means the x direction) is given by
\begin{equation}
\mathbf{j}_s =\sqrt{(j^{(2)}_{x, s_y}+j^{(2)}_{x, s_z})^2 \cos ^2 \theta + (j^{(2)}_{y, s_x})^2 \sin ^2 \theta}
\end{equation}
When the Fermi level is lower than the crossing point by 60 meV, 50 meV, 40 meV, or 30 meV, our numerical results of the spin currents as functions of $\theta$  are shown in Fig.~\ref{fig3} (c). The spin currents show giant anisotropy in the real space. They increase due to enlarged electron concentration when the Fermi level moves upwards from the conduction and edge.
To show the concentration dependence of the maximal spin currents per electron, $j_{sn}$, we present in Fig.~\ref{fig3} (d) the normalized maximal spin currents as functions of electron concentration for the two cases of the 10\% strain along the x and y directions. Here the definition of normalized maximal spin currents is similar to that of normalized charge current. Fig.~\ref{fig3} (d) indicates that the different orientations of uniaxial stress, even under the same electric field (along the x direction), will lead to very different normalized maximal spin currents.

The lifetime $\tau$ in the BiTeI bulk measured by experiment is 3.9$\times$10$^{-14}$ s~\cite{PhysRevB.87.041104} and then the relation time $\tau$ in the monolayer is assumed as this value. Assuming that the Fermi level is 60 meV lower than the crossing point  and the electric field is 1 V $\mu$m$^{-1}$ along the x direction, the ratio of the spin current to the charge current is equivalent to 0.09, and the spin current is 1.28$\times$10$^{9}$ \AA{}$^{-1}$s$^{-1}$ for the x-strain 10\%.

\subsection{Further discussion for realization}

It is useful to show the relationship between the Fermi lvel and the electron concentration. In Table~\ref{table1} we present the corresponding electron concentrations for the two 10\% strains when the Fermi level is lower than the crossing point by 30 meV, 40 meV, 50 meV, and 60 meV.
Because doped carrier concentration 10$^{13}$ cm$^{-2}$ have been achieved by back-gate gating in transition metal dichalcogenide monolayers~\cite{mak2013tightly,zhang2014electrically} and 10$^{14}$ cm$^{-2}$ by ion liquid gating in graphene~\cite{PhysRevLett.105.256805,Ye13002}, the electron concentrations ranging from 0.20 to 1.20 $\times10^{11}$ cm$^{-2}$ should be experimentally accessible.

\begin{table}[!htbp]
\caption{\label{table1} The electron concentrations (10$^{11}$ cm$^{-2}$) for different values of the Fermi level $\mathcal{E}_F$ in the presence of 10\% x-strain or y-strain due to uniaxial stress. The conduction band edge is at -69.6 meV below the crossing point for 10\% x-strain, or -67.7 meV for 10\% y-strain.}
\begin{ruledtabular}
\begin{tabular}{ccccc}
 $\mathcal{E}_F$        & -30 meV   & -40 meV      &   -50 meV  &   -60 meV   \\ \hline
 10\% x-strain             & 1.20       &  0.86             & 0.56        & 0.27\\
 10\% y-strain             & 1.02       &  0.73            & 0.46        & 0.20\\
\end{tabular}
\end{ruledtabular}
\end{table}

\begin{figure}[!htbp]
\includegraphics[width=0.48\textwidth]{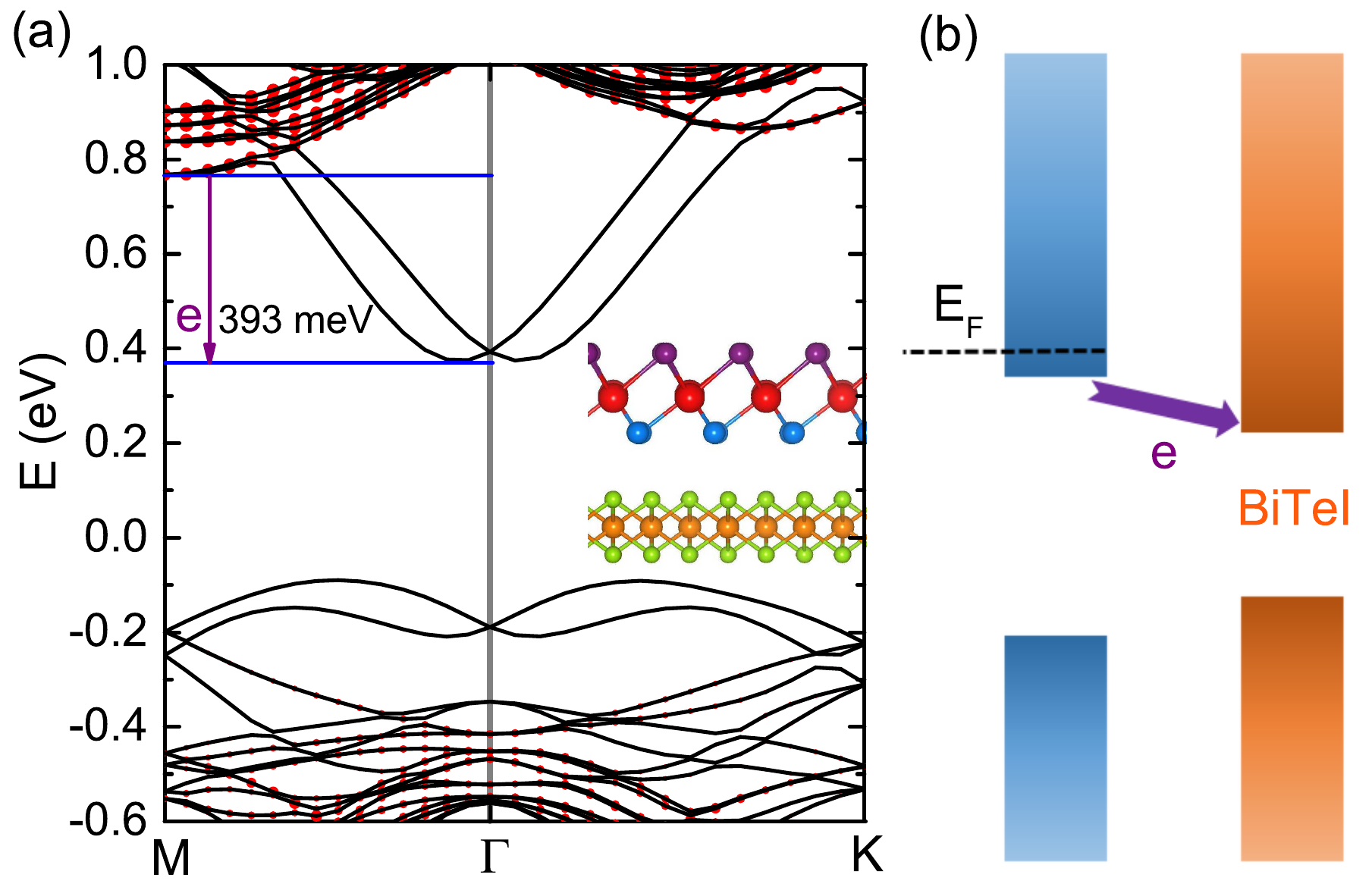}
\caption{~\label{fig4} The energy bands, with the red circles indicating the contribution of Pt atom (a), and a demonstration of the band off-sets, showing the electron transferring from the conduction edge of doped PtSe$_2$ monolayer to that of BiTeI monolayer (b) of the BiTeI/PtSe$_2$ van der Waals heterostructure. Shown in the insert in (a) is the structure of the heterostructure.}
\end{figure}

For such 2D materials as the BiTeI monolayer, the electron doping can be achieved by inter-layer charge transferring through van der Waals heterostructure formed with appropriate 2D semiconductors. PtSe$_2$ monolayer is good because it has been synthesised by molten-salt-assisted chemical vapour deposition in the experiment~\cite{Zhou2018}. We can choose the 3$\times$3 supercell of BiTeI monolayer and 2$\sqrt{3}$$\times$2$\sqrt{3}$ supercell of PtSe$_2$ monolayer to model the heterostructure. The lattice mismatch of the BiTeI/PtSe$_2$ heterostructure is 3\%. The side view of the heterostructure and the calculated energy bands are shown in Fig.~\ref{fig4}. The conduction band edge of the PtSe$_2$ monolayer is higher than that of the BiTeI monolayer, and the chemically-doped electrons in the PtSe$_2$ monolayer can easily transfer to the BiTeI monolayer. In such ways, the BiTeI monolayer can be doped with electrons, without changing its structure and energy bands.

\section{Conclusion}

In summary, we use uniaxial stress to manipulate the BiTeI monolayer semiconductor through first-principles calculations and theoretical analyses. Because the uniaxial stress destroys the $C_{3v}$ symmetry, the monolayer has only $\mathcal{M}_x$ mirror symmetry, which allows emergence of $\alpha _3k_x\sigma _z$ term in the effective model. We obtain effective anisotropic Rashba Hamiltonian through fitting the Rashba bands with the lowest conduction bands near the $\Gamma$ point up to the stress of 1.71 N/m for the x axis or 1.82 N/m for the y axis.
We find the out-of-plane spin component  in the stretched BiTeI monolayer, in addition to usual in-plane spin components in isotropic Rashba model. When electron carriers of low concentration are introduced into the monolayer, an in-plane electric field can induce the first-order charge current and second-order spin currents including a $s_z$ spin current increasing with the uniaxial stress. Such electron carriers can be realized through electron transferring from n-type doped monolayer to the BiTeI monolayer in appropriate van der Waals heterostructures, electric gating, or chemical doping. These make us believe that uniaxial stress and electric field together can open the door for controllable spintronic applications on the basis of good experimentally-realizable 2D materials such as the BiTeI monolayer.

\begin{acknowledgments}
This work is supported by the Nature Science Foundation of China (No.11574366), by the Strategic Priority Research Program of the Chinese Academy of Sciences (Grant No.XDB07000000), and by the Department of Science and Technology of China (Grant No.2016YFA0300701). The calculations were performed in the supercomputer system at Institute of Physics and the Milky Way \#2 supercomputer system at the National Supercomputer Center of Guangzhou, Guangzhou, China.
\end{acknowledgments}


\begin{thebibliography}{54}%
\makeatletter
\providecommand \@ifxundefined [1]{%
 \@ifx{#1\undefined}
}%
\providecommand \@ifnum [1]{%
 \ifnum #1\expandafter \@firstoftwo
 \else \expandafter \@secondoftwo
 \fi
}%
\providecommand \@ifx [1]{%
 \ifx #1\expandafter \@firstoftwo
 \else \expandafter \@secondoftwo
 \fi
}%
\providecommand \natexlab [1]{#1}%
\providecommand \enquote  [1]{``#1''}%
\providecommand \bibnamefont  [1]{#1}%
\providecommand \bibfnamefont [1]{#1}%
\providecommand \citenamefont [1]{#1}%
\providecommand \href@noop [0]{\@secondoftwo}%
\providecommand \href [0]{\begingroup \@sanitize@url \@href}%
\providecommand \@href[1]{\@@startlink{#1}\@@href}%
\providecommand \@@href[1]{\endgroup#1\@@endlink}%
\providecommand \@sanitize@url [0]{\catcode `\\12\catcode `\$12\catcode
  `\&12\catcode `\#12\catcode `\^12\catcode `\_12\catcode `\%12\relax}%
\providecommand \@@startlink[1]{}%
\providecommand \@@endlink[0]{}%
\providecommand \url  [0]{\begingroup\@sanitize@url \@url }%
\providecommand \@url [1]{\endgroup\@href {#1}{\urlprefix }}%
\providecommand \urlprefix  [0]{URL }%
\providecommand \Eprint [0]{\href }%
\providecommand \doibase [0]{http://dx.doi.org/}%
\providecommand \selectlanguage [0]{\@gobble}%
\providecommand \bibinfo  [0]{\@secondoftwo}%
\providecommand \bibfield  [0]{\@secondoftwo}%
\providecommand \translation [1]{[#1]}%
\providecommand \BibitemOpen [0]{}%
\providecommand \bibitemStop [0]{}%
\providecommand \bibitemNoStop [0]{.\EOS\space}%
\providecommand \EOS [0]{\spacefactor3000\relax}%
\providecommand \BibitemShut  [1]{\csname bibitem#1\endcsname}%
\let\auto@bib@innerbib\@empty
\bibitem [{\citenamefont {Rashba}\ and\ \citenamefont
  {Sheka}(1959)}]{Rashba1959}%
  \BibitemOpen
  \bibfield  {author} {\bibinfo {author} {\bibfnamefont {E.~I.}\ \bibnamefont
  {Rashba}}\ and\ \bibinfo {author} {\bibfnamefont {V.~I.}\ \bibnamefont
  {Sheka}},\ }\href@noop {} {\bibfield  {journal} {\bibinfo  {journal} {Fiz.
  Tverd. Tela.: Collected Papers}\ }\textbf {\bibinfo {volume} {2}},\ \bibinfo
  {pages} {162} (\bibinfo {year} {1959})}\BibitemShut {NoStop}%
\bibitem [{\citenamefont {Rashba}(1960)}]{Rashba1960}%
  \BibitemOpen
  \bibfield  {author} {\bibinfo {author} {\bibfnamefont {E.~I.}\ \bibnamefont
  {Rashba}},\ }\href@noop {} {\bibfield  {journal} {\bibinfo  {journal} {Fiz.
  Tverd. Tela.}\ }\textbf {\bibinfo {volume} {2}},\ \bibinfo {pages} {1224}
  (\bibinfo {year} {1960})}\BibitemShut {NoStop}%
\bibitem [{\citenamefont {Dyakonov~(ed.)}(2017)}]{SpinSec2017}%
  \BibitemOpen
  \bibfield  {author} {\bibinfo {author} {\bibfnamefont {M.~I.}\ \bibnamefont
  {Dyakonov~(ed.)}},\ }\href {\doibase 10.1007/978-3-319-65436-2_1} {\emph
  {\bibinfo {title} {Spin Physics in Semiconductors}}},\ \bibinfo {series}
  {Springer Series in Solid-State Sciences}, Vol.\ \bibinfo {volume} {157}\
  (\bibinfo  {publisher} {Springer International Publishing AG},\ \bibinfo
  {year} {2017})\BibitemShut {NoStop}%
\bibitem [{\citenamefont {Bychkov}\ and\ \citenamefont
  {Rashba}(1984)}]{bychkov1984properties}%
  \BibitemOpen
  \bibfield  {author} {\bibinfo {author} {\bibfnamefont {Y.~A.}\ \bibnamefont
  {Bychkov}}\ and\ \bibinfo {author} {\bibfnamefont {{\'E}.~I.}\ \bibnamefont
  {Rashba}},\ }\href@noop {} {\bibfield  {journal} {\bibinfo  {journal} {JETP
  Lett.}\ }\textbf {\bibinfo {volume} {39}},\ \bibinfo {pages} {78} (\bibinfo
  {year} {1984})}\BibitemShut {NoStop}%
\bibitem [{\citenamefont {Manchon}\ \emph {et~al.}(2015)\citenamefont
  {Manchon}, \citenamefont {Koo}, \citenamefont {Nitta}, \citenamefont
  {Frolov},\ and\ \citenamefont {Duine}}]{Manchon2015}%
  \BibitemOpen
  \bibfield  {author} {\bibinfo {author} {\bibfnamefont {A.}~\bibnamefont
  {Manchon}}, \bibinfo {author} {\bibfnamefont {H.~C.}\ \bibnamefont {Koo}},
  \bibinfo {author} {\bibfnamefont {J.}~\bibnamefont {Nitta}}, \bibinfo
  {author} {\bibfnamefont {S.~M.}\ \bibnamefont {Frolov}}, \ and\ \bibinfo
  {author} {\bibfnamefont {R.~A.}\ \bibnamefont {Duine}},\ }\href
  {https://doi.org/10.1038/nmat4360} {\bibfield  {journal} {\bibinfo  {journal}
  {Nat. Mater.}\ }\textbf {\bibinfo {volume} {14}},\ \bibinfo {pages} {871}
  (\bibinfo {year} {2015})}\BibitemShut {NoStop}%
\bibitem [{\citenamefont {Hu}\ \emph {et~al.}(2018)\citenamefont {Hu},
  \citenamefont {Jia}, \citenamefont {Zhao}, \citenamefont {Wu}, \citenamefont
  {Stroppa},\ and\ \citenamefont {Ren}}]{PhysRevB.97.235404}%
  \BibitemOpen
  \bibfield  {author} {\bibinfo {author} {\bibfnamefont {T.}~\bibnamefont
  {Hu}}, \bibinfo {author} {\bibfnamefont {F.}~\bibnamefont {Jia}}, \bibinfo
  {author} {\bibfnamefont {G.}~\bibnamefont {Zhao}}, \bibinfo {author}
  {\bibfnamefont {J.}~\bibnamefont {Wu}}, \bibinfo {author} {\bibfnamefont
  {A.}~\bibnamefont {Stroppa}}, \ and\ \bibinfo {author} {\bibfnamefont
  {W.}~\bibnamefont {Ren}},\ }\href {\doibase 10.1103/PhysRevB.97.235404}
  {\bibfield  {journal} {\bibinfo  {journal} {Phys. Rev. B}\ }\textbf {\bibinfo
  {volume} {97}},\ \bibinfo {pages} {235404} (\bibinfo {year}
  {2018})}\BibitemShut {NoStop}%
\bibitem [{\citenamefont {Singh}\ and\ \citenamefont
  {Romero}(2017)}]{PhysRevB.95.165444}%
  \BibitemOpen
  \bibfield  {author} {\bibinfo {author} {\bibfnamefont {S.}~\bibnamefont
  {Singh}}\ and\ \bibinfo {author} {\bibfnamefont {A.~H.}\ \bibnamefont
  {Romero}},\ }\href {\doibase 10.1103/PhysRevB.95.165444} {\bibfield
  {journal} {\bibinfo  {journal} {Phys. Rev. B}\ }\textbf {\bibinfo {volume}
  {95}},\ \bibinfo {pages} {165444} (\bibinfo {year} {2017})}\BibitemShut
  {NoStop}%
\bibitem [{\citenamefont {Yuan}\ \emph {et~al.}(2018)\citenamefont {Yuan},
  \citenamefont {Cai}, \citenamefont {Shen}, \citenamefont {Xiao},
  \citenamefont {Ren}, \citenamefont {Wang}, \citenamefont {Feng},\ and\
  \citenamefont {Yan}}]{YUAN2018163}%
  \BibitemOpen
  \bibfield  {author} {\bibinfo {author} {\bibfnamefont {J.}~\bibnamefont
  {Yuan}}, \bibinfo {author} {\bibfnamefont {Y.}~\bibnamefont {Cai}}, \bibinfo
  {author} {\bibfnamefont {L.}~\bibnamefont {Shen}}, \bibinfo {author}
  {\bibfnamefont {Y.}~\bibnamefont {Xiao}}, \bibinfo {author} {\bibfnamefont
  {J.-C.}\ \bibnamefont {Ren}}, \bibinfo {author} {\bibfnamefont
  {A.}~\bibnamefont {Wang}}, \bibinfo {author} {\bibfnamefont {Y.~P.}\
  \bibnamefont {Feng}}, \ and\ \bibinfo {author} {\bibfnamefont
  {X.}~\bibnamefont {Yan}},\ }\href {\doibase
  https://doi.org/10.1016/j.nanoen.2018.07.041} {\bibfield  {journal} {\bibinfo
   {journal} {Nano Energy}\ }\textbf {\bibinfo {volume} {52}},\ \bibinfo
  {pages} {163 } (\bibinfo {year} {2018})}\BibitemShut {NoStop}%
\bibitem [{\citenamefont {Lu}\ \emph {et~al.}(2017)\citenamefont {Lu},
  \citenamefont {Zhu}, \citenamefont {Xiao}, \citenamefont {Chuu},
  \citenamefont {Han}, \citenamefont {Chiu}, \citenamefont {Cheng},
  \citenamefont {Yang}, \citenamefont {Wei}, \citenamefont {Yang},
  \citenamefont {Wang}, \citenamefont {Sokaras}, \citenamefont {Nordlund},
  \citenamefont {Yang}, \citenamefont {Muller}, \citenamefont {Chou},
  \citenamefont {Zhang},\ and\ \citenamefont {Li}}]{Lu2017}%
  \BibitemOpen
  \bibfield  {author} {\bibinfo {author} {\bibfnamefont {A.-Y.}\ \bibnamefont
  {Lu}}, \bibinfo {author} {\bibfnamefont {H.}~\bibnamefont {Zhu}}, \bibinfo
  {author} {\bibfnamefont {J.}~\bibnamefont {Xiao}}, \bibinfo {author}
  {\bibfnamefont {C.-P.}\ \bibnamefont {Chuu}}, \bibinfo {author}
  {\bibfnamefont {Y.}~\bibnamefont {Han}}, \bibinfo {author} {\bibfnamefont
  {M.-H.}\ \bibnamefont {Chiu}}, \bibinfo {author} {\bibfnamefont {C.-C.}\
  \bibnamefont {Cheng}}, \bibinfo {author} {\bibfnamefont {C.-W.}\ \bibnamefont
  {Yang}}, \bibinfo {author} {\bibfnamefont {K.-H.}\ \bibnamefont {Wei}},
  \bibinfo {author} {\bibfnamefont {Y.}~\bibnamefont {Yang}}, \bibinfo {author}
  {\bibfnamefont {Y.}~\bibnamefont {Wang}}, \bibinfo {author} {\bibfnamefont
  {D.}~\bibnamefont {Sokaras}}, \bibinfo {author} {\bibfnamefont
  {D.}~\bibnamefont {Nordlund}}, \bibinfo {author} {\bibfnamefont
  {P.}~\bibnamefont {Yang}}, \bibinfo {author} {\bibfnamefont {D.~A.}\
  \bibnamefont {Muller}}, \bibinfo {author} {\bibfnamefont {M.-Y.}\
  \bibnamefont {Chou}}, \bibinfo {author} {\bibfnamefont {X.}~\bibnamefont
  {Zhang}}, \ and\ \bibinfo {author} {\bibfnamefont {L.-J.}\ \bibnamefont
  {Li}},\ }\href {https://doi.org/10.1038/nnano.2017.100} {\bibfield  {journal}
  {\bibinfo  {journal} {Nature Nanotechnology}\ }\textbf {\bibinfo {volume}
  {12}},\ \bibinfo {pages} {744} (\bibinfo {year} {2017})}\BibitemShut
  {NoStop}%
\bibitem [{\citenamefont {Hanakata}\ \emph {et~al.}(2017)\citenamefont
  {Hanakata}, \citenamefont {Rodin}, \citenamefont {Carvalho}, \citenamefont
  {Park}, \citenamefont {Campbell},\ and\ \citenamefont
  {Castro~Neto}}]{PhysRevB.96.161401}%
  \BibitemOpen
  \bibfield  {author} {\bibinfo {author} {\bibfnamefont {P.~Z.}\ \bibnamefont
  {Hanakata}}, \bibinfo {author} {\bibfnamefont {A.~S.}\ \bibnamefont {Rodin}},
  \bibinfo {author} {\bibfnamefont {A.}~\bibnamefont {Carvalho}}, \bibinfo
  {author} {\bibfnamefont {H.~S.}\ \bibnamefont {Park}}, \bibinfo {author}
  {\bibfnamefont {D.~K.}\ \bibnamefont {Campbell}}, \ and\ \bibinfo {author}
  {\bibfnamefont {A.~H.}\ \bibnamefont {Castro~Neto}},\ }\href {\doibase
  10.1103/PhysRevB.96.161401} {\bibfield  {journal} {\bibinfo  {journal} {Phys.
  Rev. B}\ }\textbf {\bibinfo {volume} {96}},\ \bibinfo {pages} {161401}
  (\bibinfo {year} {2017})}\BibitemShut {NoStop}%
\bibitem [{\citenamefont {Hanakata}\ \emph {et~al.}(2018)\citenamefont
  {Hanakata}, \citenamefont {Rodin}, \citenamefont {Park}, \citenamefont
  {Campbell},\ and\ \citenamefont {Castro~Neto}}]{PhysRevB.97.235312}%
  \BibitemOpen
  \bibfield  {author} {\bibinfo {author} {\bibfnamefont {P.~Z.}\ \bibnamefont
  {Hanakata}}, \bibinfo {author} {\bibfnamefont {A.~S.}\ \bibnamefont {Rodin}},
  \bibinfo {author} {\bibfnamefont {H.~S.}\ \bibnamefont {Park}}, \bibinfo
  {author} {\bibfnamefont {D.~K.}\ \bibnamefont {Campbell}}, \ and\ \bibinfo
  {author} {\bibfnamefont {A.~H.}\ \bibnamefont {Castro~Neto}},\ }\href
  {\doibase 10.1103/PhysRevB.97.235312} {\bibfield  {journal} {\bibinfo
  {journal} {Phys. Rev. B}\ }\textbf {\bibinfo {volume} {97}},\ \bibinfo
  {pages} {235312} (\bibinfo {year} {2018})}\BibitemShut {NoStop}%
\bibitem [{\citenamefont {Noor-A-Alam}\ \emph {et~al.}(2018)\citenamefont
  {Noor-A-Alam}, \citenamefont {Lee}, \citenamefont {Lee}, \citenamefont
  {Choi},\ and\ \citenamefont {Lee}}]{ag2te}%
  \BibitemOpen
  \bibfield  {author} {\bibinfo {author} {\bibfnamefont {M.}~\bibnamefont
  {Noor-A-Alam}}, \bibinfo {author} {\bibfnamefont {M.}~\bibnamefont {Lee}},
  \bibinfo {author} {\bibfnamefont {H.-J.}\ \bibnamefont {Lee}}, \bibinfo
  {author} {\bibfnamefont {K.}~\bibnamefont {Choi}}, \ and\ \bibinfo {author}
  {\bibfnamefont {J.~H.}\ \bibnamefont {Lee}},\ }\href
  {http://stacks.iop.org/0953-8984/30/i=38/a=385502} {\bibfield  {journal}
  {\bibinfo  {journal} {J. Phys. Condens. Mat.}\ }\textbf {\bibinfo {volume}
  {30}},\ \bibinfo {pages} {385502} (\bibinfo {year} {2018})}\BibitemShut
  {NoStop}%
\bibitem [{\citenamefont {Premasiri}\ \emph {et~al.}(2018)\citenamefont
  {Premasiri}, \citenamefont {Radha}, \citenamefont {Sucharitakul},
  \citenamefont {Kumar}, \citenamefont {Sankar}, \citenamefont {Chou},
  \citenamefont {Chen},\ and\ \citenamefont {Gao}}]{acs.nanolett.8b01462}%
  \BibitemOpen
  \bibfield  {author} {\bibinfo {author} {\bibfnamefont {K.}~\bibnamefont
  {Premasiri}}, \bibinfo {author} {\bibfnamefont {S.~K.}\ \bibnamefont
  {Radha}}, \bibinfo {author} {\bibfnamefont {S.}~\bibnamefont {Sucharitakul}},
  \bibinfo {author} {\bibfnamefont {U.~R.}\ \bibnamefont {Kumar}}, \bibinfo
  {author} {\bibfnamefont {R.}~\bibnamefont {Sankar}}, \bibinfo {author}
  {\bibfnamefont {F.-C.}\ \bibnamefont {Chou}}, \bibinfo {author}
  {\bibfnamefont {Y.-T.}\ \bibnamefont {Chen}}, \ and\ \bibinfo {author}
  {\bibfnamefont {X.~P.~A.}\ \bibnamefont {Gao}},\ }\href {\doibase
  10.1021/acs.nanolett.8b01462} {\bibfield  {journal} {\bibinfo  {journal}
  {Nano Lett.}\ }\textbf {\bibinfo {volume} {18}},\ \bibinfo {pages} {4403}
  (\bibinfo {year} {2018})}\BibitemShut {NoStop}%
\bibitem [{\citenamefont {Zhang}\ and\ \citenamefont
  {Schwingenschl\"ogl}(2018)}]{PhysRevB.97.155415}%
  \BibitemOpen
  \bibfield  {author} {\bibinfo {author} {\bibfnamefont {Q.}~\bibnamefont
  {Zhang}}\ and\ \bibinfo {author} {\bibfnamefont {U.}~\bibnamefont
  {Schwingenschl\"ogl}},\ }\href {\doibase 10.1103/PhysRevB.97.155415}
  {\bibfield  {journal} {\bibinfo  {journal} {Phys. Rev. B}\ }\textbf {\bibinfo
  {volume} {97}},\ \bibinfo {pages} {155415} (\bibinfo {year}
  {2018})}\BibitemShut {NoStop}%
\bibitem [{\citenamefont {Ishizaka}\ \emph {et~al.}(2011)\citenamefont
  {Ishizaka}, \citenamefont {Bahramy}, \citenamefont {Murakawa}, \citenamefont
  {Sakano}, \citenamefont {Shimojima}, \citenamefont {Sonobe}, \citenamefont
  {Koizumi}, \citenamefont {Shin}, \citenamefont {Miyahara}, \citenamefont
  {Kimura}, \citenamefont {Miyamoto}, \citenamefont {Okuda}, \citenamefont
  {Namatame}, \citenamefont {Taniguchi}, \citenamefont {Arita}, \citenamefont
  {Nagaosa}, \citenamefont {Kobayashi}, \citenamefont {Murakami}, \citenamefont
  {Kumai}, \citenamefont {Kaneko}, \citenamefont {Onose},\ and\ \citenamefont
  {Tokura}}]{Ishizaka2011}%
  \BibitemOpen
  \bibfield  {author} {\bibinfo {author} {\bibfnamefont {K.}~\bibnamefont
  {Ishizaka}}, \bibinfo {author} {\bibfnamefont {M.~S.}\ \bibnamefont
  {Bahramy}}, \bibinfo {author} {\bibfnamefont {H.}~\bibnamefont {Murakawa}},
  \bibinfo {author} {\bibfnamefont {M.}~\bibnamefont {Sakano}}, \bibinfo
  {author} {\bibfnamefont {T.}~\bibnamefont {Shimojima}}, \bibinfo {author}
  {\bibfnamefont {T.}~\bibnamefont {Sonobe}}, \bibinfo {author} {\bibfnamefont
  {K.}~\bibnamefont {Koizumi}}, \bibinfo {author} {\bibfnamefont
  {S.}~\bibnamefont {Shin}}, \bibinfo {author} {\bibfnamefont {H.}~\bibnamefont
  {Miyahara}}, \bibinfo {author} {\bibfnamefont {A.}~\bibnamefont {Kimura}},
  \bibinfo {author} {\bibfnamefont {K.}~\bibnamefont {Miyamoto}}, \bibinfo
  {author} {\bibfnamefont {T.}~\bibnamefont {Okuda}}, \bibinfo {author}
  {\bibfnamefont {H.}~\bibnamefont {Namatame}}, \bibinfo {author}
  {\bibfnamefont {M.}~\bibnamefont {Taniguchi}}, \bibinfo {author}
  {\bibfnamefont {R.}~\bibnamefont {Arita}}, \bibinfo {author} {\bibfnamefont
  {N.}~\bibnamefont {Nagaosa}}, \bibinfo {author} {\bibfnamefont
  {K.}~\bibnamefont {Kobayashi}}, \bibinfo {author} {\bibfnamefont
  {Y.}~\bibnamefont {Murakami}}, \bibinfo {author} {\bibfnamefont
  {R.}~\bibnamefont {Kumai}}, \bibinfo {author} {\bibfnamefont
  {Y.}~\bibnamefont {Kaneko}}, \bibinfo {author} {\bibfnamefont
  {Y.}~\bibnamefont {Onose}}, \ and\ \bibinfo {author} {\bibfnamefont
  {Y.}~\bibnamefont {Tokura}},\ }\href {https://doi.org/10.1038/nmat3051}
  {\bibfield  {journal} {\bibinfo  {journal} {Nat. Mater.}\ }\textbf {\bibinfo
  {volume} {10}},\ \bibinfo {pages} {521} (\bibinfo {year} {2011})}\BibitemShut
  {NoStop}%
\bibitem [{\citenamefont {Crepaldi}\ \emph {et~al.}(2012)\citenamefont
  {Crepaldi}, \citenamefont {Moreschini}, \citenamefont {Aut\`es},
  \citenamefont {Tournier-Colletta}, \citenamefont {Moser}, \citenamefont
  {Virk}, \citenamefont {Berger}, \citenamefont {Bugnon}, \citenamefont
  {Chang}, \citenamefont {Kern}, \citenamefont {Bostwick}, \citenamefont
  {Rotenberg}, \citenamefont {Yazyev},\ and\ \citenamefont
  {Grioni}}]{PhysRevLett.109.096803}%
  \BibitemOpen
  \bibfield  {author} {\bibinfo {author} {\bibfnamefont {A.}~\bibnamefont
  {Crepaldi}}, \bibinfo {author} {\bibfnamefont {L.}~\bibnamefont
  {Moreschini}}, \bibinfo {author} {\bibfnamefont {G.}~\bibnamefont {Aut\`es}},
  \bibinfo {author} {\bibfnamefont {C.}~\bibnamefont {Tournier-Colletta}},
  \bibinfo {author} {\bibfnamefont {S.}~\bibnamefont {Moser}}, \bibinfo
  {author} {\bibfnamefont {N.}~\bibnamefont {Virk}}, \bibinfo {author}
  {\bibfnamefont {H.}~\bibnamefont {Berger}}, \bibinfo {author} {\bibfnamefont
  {P.}~\bibnamefont {Bugnon}}, \bibinfo {author} {\bibfnamefont {Y.~J.}\
  \bibnamefont {Chang}}, \bibinfo {author} {\bibfnamefont {K.}~\bibnamefont
  {Kern}}, \bibinfo {author} {\bibfnamefont {A.}~\bibnamefont {Bostwick}},
  \bibinfo {author} {\bibfnamefont {E.}~\bibnamefont {Rotenberg}}, \bibinfo
  {author} {\bibfnamefont {O.~V.}\ \bibnamefont {Yazyev}}, \ and\ \bibinfo
  {author} {\bibfnamefont {M.}~\bibnamefont {Grioni}},\ }\href {\doibase
  10.1103/PhysRevLett.109.096803} {\bibfield  {journal} {\bibinfo  {journal}
  {Phys. Rev. Lett.}\ }\textbf {\bibinfo {volume} {109}},\ \bibinfo {pages}
  {096803} (\bibinfo {year} {2012})}\BibitemShut {NoStop}%
\bibitem [{\citenamefont {Landolt}\ \emph {et~al.}(2012)\citenamefont
  {Landolt}, \citenamefont {Eremeev}, \citenamefont {Koroteev}, \citenamefont
  {Slomski}, \citenamefont {Muff}, \citenamefont {Neupert}, \citenamefont
  {Kobayashi}, \citenamefont {Strocov}, \citenamefont {Schmitt}, \citenamefont
  {Aliev}, \citenamefont {Babanly}, \citenamefont {Amiraslanov}, \citenamefont
  {Chulkov}, \citenamefont {Osterwalder},\ and\ \citenamefont
  {Dil}}]{PhysRevLett.109.116403}%
  \BibitemOpen
  \bibfield  {author} {\bibinfo {author} {\bibfnamefont {G.}~\bibnamefont
  {Landolt}}, \bibinfo {author} {\bibfnamefont {S.~V.}\ \bibnamefont
  {Eremeev}}, \bibinfo {author} {\bibfnamefont {Y.~M.}\ \bibnamefont
  {Koroteev}}, \bibinfo {author} {\bibfnamefont {B.}~\bibnamefont {Slomski}},
  \bibinfo {author} {\bibfnamefont {S.}~\bibnamefont {Muff}}, \bibinfo {author}
  {\bibfnamefont {T.}~\bibnamefont {Neupert}}, \bibinfo {author} {\bibfnamefont
  {M.}~\bibnamefont {Kobayashi}}, \bibinfo {author} {\bibfnamefont {V.~N.}\
  \bibnamefont {Strocov}}, \bibinfo {author} {\bibfnamefont {T.}~\bibnamefont
  {Schmitt}}, \bibinfo {author} {\bibfnamefont {Z.~S.}\ \bibnamefont {Aliev}},
  \bibinfo {author} {\bibfnamefont {M.~B.}\ \bibnamefont {Babanly}}, \bibinfo
  {author} {\bibfnamefont {I.~R.}\ \bibnamefont {Amiraslanov}}, \bibinfo
  {author} {\bibfnamefont {E.~V.}\ \bibnamefont {Chulkov}}, \bibinfo {author}
  {\bibfnamefont {J.}~\bibnamefont {Osterwalder}}, \ and\ \bibinfo {author}
  {\bibfnamefont {J.~H.}\ \bibnamefont {Dil}},\ }\href {\doibase
  10.1103/PhysRevLett.109.116403} {\bibfield  {journal} {\bibinfo  {journal}
  {Phys. Rev. Lett.}\ }\textbf {\bibinfo {volume} {109}},\ \bibinfo {pages}
  {116403} (\bibinfo {year} {2012})}\BibitemShut {NoStop}%
\bibitem [{\citenamefont {Eremeev}\ \emph {et~al.}(2012)\citenamefont
  {Eremeev}, \citenamefont {Nechaev}, \citenamefont {Koroteev}, \citenamefont
  {Echenique},\ and\ \citenamefont {Chulkov}}]{PhysRevLett.108.246802}%
  \BibitemOpen
  \bibfield  {author} {\bibinfo {author} {\bibfnamefont {S.~V.}\ \bibnamefont
  {Eremeev}}, \bibinfo {author} {\bibfnamefont {I.~A.}\ \bibnamefont
  {Nechaev}}, \bibinfo {author} {\bibfnamefont {Y.~M.}\ \bibnamefont
  {Koroteev}}, \bibinfo {author} {\bibfnamefont {P.~M.}\ \bibnamefont
  {Echenique}}, \ and\ \bibinfo {author} {\bibfnamefont {E.~V.}\ \bibnamefont
  {Chulkov}},\ }\href {\doibase 10.1103/PhysRevLett.108.246802} {\bibfield
  {journal} {\bibinfo  {journal} {Phys. Rev. Lett.}\ }\textbf {\bibinfo
  {volume} {108}},\ \bibinfo {pages} {246802} (\bibinfo {year}
  {2012})}\BibitemShut {NoStop}%
\bibitem [{\citenamefont {Sakano}\ \emph {et~al.}(2013)\citenamefont {Sakano},
  \citenamefont {Bahramy}, \citenamefont {Katayama}, \citenamefont {Shimojima},
  \citenamefont {Murakawa}, \citenamefont {Kaneko}, \citenamefont {Malaeb},
  \citenamefont {Shin}, \citenamefont {Ono}, \citenamefont {Kumigashira},
  \citenamefont {Arita}, \citenamefont {Nagaosa}, \citenamefont {Hwang},
  \citenamefont {Tokura},\ and\ \citenamefont
  {Ishizaka}}]{PhysRevLett.110.107204}%
  \BibitemOpen
  \bibfield  {author} {\bibinfo {author} {\bibfnamefont {M.}~\bibnamefont
  {Sakano}}, \bibinfo {author} {\bibfnamefont {M.~S.}\ \bibnamefont {Bahramy}},
  \bibinfo {author} {\bibfnamefont {A.}~\bibnamefont {Katayama}}, \bibinfo
  {author} {\bibfnamefont {T.}~\bibnamefont {Shimojima}}, \bibinfo {author}
  {\bibfnamefont {H.}~\bibnamefont {Murakawa}}, \bibinfo {author}
  {\bibfnamefont {Y.}~\bibnamefont {Kaneko}}, \bibinfo {author} {\bibfnamefont
  {W.}~\bibnamefont {Malaeb}}, \bibinfo {author} {\bibfnamefont
  {S.}~\bibnamefont {Shin}}, \bibinfo {author} {\bibfnamefont {K.}~\bibnamefont
  {Ono}}, \bibinfo {author} {\bibfnamefont {H.}~\bibnamefont {Kumigashira}},
  \bibinfo {author} {\bibfnamefont {R.}~\bibnamefont {Arita}}, \bibinfo
  {author} {\bibfnamefont {N.}~\bibnamefont {Nagaosa}}, \bibinfo {author}
  {\bibfnamefont {H.~Y.}\ \bibnamefont {Hwang}}, \bibinfo {author}
  {\bibfnamefont {Y.}~\bibnamefont {Tokura}}, \ and\ \bibinfo {author}
  {\bibfnamefont {K.}~\bibnamefont {Ishizaka}},\ }\href {\doibase
  10.1103/PhysRevLett.110.107204} {\bibfield  {journal} {\bibinfo  {journal}
  {Phys. Rev. Lett.}\ }\textbf {\bibinfo {volume} {110}},\ \bibinfo {pages}
  {107204} (\bibinfo {year} {2013})}\BibitemShut {NoStop}%
\bibitem [{\citenamefont {Monserrat}\ and\ \citenamefont
  {Vanderbilt}(2017)}]{PhysRevMaterials.1.054201}%
  \BibitemOpen
  \bibfield  {author} {\bibinfo {author} {\bibfnamefont {B.}~\bibnamefont
  {Monserrat}}\ and\ \bibinfo {author} {\bibfnamefont {D.}~\bibnamefont
  {Vanderbilt}},\ }\href {\doibase 10.1103/PhysRevMaterials.1.054201}
  {\bibfield  {journal} {\bibinfo  {journal} {Phys. Rev. Mater.}\ }\textbf
  {\bibinfo {volume} {1}},\ \bibinfo {pages} {054201} (\bibinfo {year}
  {2017})}\BibitemShut {NoStop}%
\bibitem [{\citenamefont {Bord\'acs}\ \emph {et~al.}(2013)\citenamefont
  {Bord\'acs}, \citenamefont {Checkelsky}, \citenamefont {Murakawa},
  \citenamefont {Hwang},\ and\ \citenamefont
  {Tokura}}]{PhysRevLett.111.166403}%
  \BibitemOpen
  \bibfield  {author} {\bibinfo {author} {\bibfnamefont {S.}~\bibnamefont
  {Bord\'acs}}, \bibinfo {author} {\bibfnamefont {J.~G.}\ \bibnamefont
  {Checkelsky}}, \bibinfo {author} {\bibfnamefont {H.}~\bibnamefont
  {Murakawa}}, \bibinfo {author} {\bibfnamefont {H.~Y.}\ \bibnamefont {Hwang}},
  \ and\ \bibinfo {author} {\bibfnamefont {Y.}~\bibnamefont {Tokura}},\ }\href
  {\doibase 10.1103/PhysRevLett.111.166403} {\bibfield  {journal} {\bibinfo
  {journal} {Phys. Rev. Lett.}\ }\textbf {\bibinfo {volume} {111}},\ \bibinfo
  {pages} {166403} (\bibinfo {year} {2013})}\BibitemShut {NoStop}%
\bibitem [{\citenamefont {Sakano}\ \emph {et~al.}(2012)\citenamefont {Sakano},
  \citenamefont {Miyawaki}, \citenamefont {Chainani}, \citenamefont {Takata},
  \citenamefont {Sonobe}, \citenamefont {Shimojima}, \citenamefont {Oura},
  \citenamefont {Shin}, \citenamefont {Bahramy}, \citenamefont {Arita},
  \citenamefont {Nagaosa}, \citenamefont {Murakawa}, \citenamefont {Kaneko},
  \citenamefont {Tokura},\ and\ \citenamefont {Ishizaka}}]{PhysRevB.86.085204}%
  \BibitemOpen
  \bibfield  {author} {\bibinfo {author} {\bibfnamefont {M.}~\bibnamefont
  {Sakano}}, \bibinfo {author} {\bibfnamefont {J.}~\bibnamefont {Miyawaki}},
  \bibinfo {author} {\bibfnamefont {A.}~\bibnamefont {Chainani}}, \bibinfo
  {author} {\bibfnamefont {Y.}~\bibnamefont {Takata}}, \bibinfo {author}
  {\bibfnamefont {T.}~\bibnamefont {Sonobe}}, \bibinfo {author} {\bibfnamefont
  {T.}~\bibnamefont {Shimojima}}, \bibinfo {author} {\bibfnamefont
  {M.}~\bibnamefont {Oura}}, \bibinfo {author} {\bibfnamefont {S.}~\bibnamefont
  {Shin}}, \bibinfo {author} {\bibfnamefont {M.~S.}\ \bibnamefont {Bahramy}},
  \bibinfo {author} {\bibfnamefont {R.}~\bibnamefont {Arita}}, \bibinfo
  {author} {\bibfnamefont {N.}~\bibnamefont {Nagaosa}}, \bibinfo {author}
  {\bibfnamefont {H.}~\bibnamefont {Murakawa}}, \bibinfo {author}
  {\bibfnamefont {Y.}~\bibnamefont {Kaneko}}, \bibinfo {author} {\bibfnamefont
  {Y.}~\bibnamefont {Tokura}}, \ and\ \bibinfo {author} {\bibfnamefont
  {K.}~\bibnamefont {Ishizaka}},\ }\href {\doibase 10.1103/PhysRevB.86.085204}
  {\bibfield  {journal} {\bibinfo  {journal} {Phys. Rev. B}\ }\textbf {\bibinfo
  {volume} {86}},\ \bibinfo {pages} {085204} (\bibinfo {year}
  {2012})}\BibitemShut {NoStop}%
\bibitem [{\citenamefont {Butler}\ \emph {et~al.}(2014)\citenamefont {Butler},
  \citenamefont {Yang}, \citenamefont {Hong}, \citenamefont {Hsu},
  \citenamefont {Sankar}, \citenamefont {Lu}, \citenamefont {Lu}, \citenamefont
  {Yang}, \citenamefont {Shiu}, \citenamefont {Chen}, \citenamefont {Kaun},
  \citenamefont {Shu}, \citenamefont {Chou},\ and\ \citenamefont
  {Lin}}]{Butler2014}%
  \BibitemOpen
  \bibfield  {author} {\bibinfo {author} {\bibfnamefont {C.~J.}\ \bibnamefont
  {Butler}}, \bibinfo {author} {\bibfnamefont {H.-H.}\ \bibnamefont {Yang}},
  \bibinfo {author} {\bibfnamefont {J.-Y.}\ \bibnamefont {Hong}}, \bibinfo
  {author} {\bibfnamefont {S.-H.}\ \bibnamefont {Hsu}}, \bibinfo {author}
  {\bibfnamefont {R.}~\bibnamefont {Sankar}}, \bibinfo {author} {\bibfnamefont
  {C.-I.}\ \bibnamefont {Lu}}, \bibinfo {author} {\bibfnamefont {H.-Y.}\
  \bibnamefont {Lu}}, \bibinfo {author} {\bibfnamefont {K.-H.~O.}\ \bibnamefont
  {Yang}}, \bibinfo {author} {\bibfnamefont {H.-W.}\ \bibnamefont {Shiu}},
  \bibinfo {author} {\bibfnamefont {C.-H.}\ \bibnamefont {Chen}}, \bibinfo
  {author} {\bibfnamefont {C.-C.}\ \bibnamefont {Kaun}}, \bibinfo {author}
  {\bibfnamefont {G.-J.}\ \bibnamefont {Shu}}, \bibinfo {author} {\bibfnamefont
  {F.-C.}\ \bibnamefont {Chou}}, \ and\ \bibinfo {author} {\bibfnamefont
  {M.-T.}\ \bibnamefont {Lin}},\ }\href {https://doi.org/10.1038/ncomms5066}
  {\bibfield  {journal} {\bibinfo  {journal} {Nat. Commun.}\ }\textbf {\bibinfo
  {volume} {5}},\ \bibinfo {pages} {4066} (\bibinfo {year} {2014})}\BibitemShut
  {NoStop}%
\bibitem [{\citenamefont {Kohsaka}\ \emph {et~al.}(2015)\citenamefont
  {Kohsaka}, \citenamefont {Kanou}, \citenamefont {Takagi}, \citenamefont
  {Hanaguri},\ and\ \citenamefont {Sasagawa}}]{PhysRevB.91.245312}%
  \BibitemOpen
  \bibfield  {author} {\bibinfo {author} {\bibfnamefont {Y.}~\bibnamefont
  {Kohsaka}}, \bibinfo {author} {\bibfnamefont {M.}~\bibnamefont {Kanou}},
  \bibinfo {author} {\bibfnamefont {H.}~\bibnamefont {Takagi}}, \bibinfo
  {author} {\bibfnamefont {T.}~\bibnamefont {Hanaguri}}, \ and\ \bibinfo
  {author} {\bibfnamefont {T.}~\bibnamefont {Sasagawa}},\ }\href {\doibase
  10.1103/PhysRevB.91.245312} {\bibfield  {journal} {\bibinfo  {journal} {Phys.
  Rev. B}\ }\textbf {\bibinfo {volume} {91}},\ \bibinfo {pages} {245312}
  (\bibinfo {year} {2015})}\BibitemShut {NoStop}%
\bibitem [{\citenamefont {Lee}\ \emph {et~al.}(2011)\citenamefont {Lee},
  \citenamefont {Schober}, \citenamefont {Bahramy}, \citenamefont {Murakawa},
  \citenamefont {Onose}, \citenamefont {Arita}, \citenamefont {Nagaosa},\ and\
  \citenamefont {Tokura}}]{PhysRevLett.107.117401}%
  \BibitemOpen
  \bibfield  {author} {\bibinfo {author} {\bibfnamefont {J.~S.}\ \bibnamefont
  {Lee}}, \bibinfo {author} {\bibfnamefont {G.~A.~H.}\ \bibnamefont {Schober}},
  \bibinfo {author} {\bibfnamefont {M.~S.}\ \bibnamefont {Bahramy}}, \bibinfo
  {author} {\bibfnamefont {H.}~\bibnamefont {Murakawa}}, \bibinfo {author}
  {\bibfnamefont {Y.}~\bibnamefont {Onose}}, \bibinfo {author} {\bibfnamefont
  {R.}~\bibnamefont {Arita}}, \bibinfo {author} {\bibfnamefont
  {N.}~\bibnamefont {Nagaosa}}, \ and\ \bibinfo {author} {\bibfnamefont
  {Y.}~\bibnamefont {Tokura}},\ }\href {\doibase
  10.1103/PhysRevLett.107.117401} {\bibfield  {journal} {\bibinfo  {journal}
  {Phys. Rev. Lett.}\ }\textbf {\bibinfo {volume} {107}},\ \bibinfo {pages}
  {117401} (\bibinfo {year} {2011})}\BibitemShut {NoStop}%
\bibitem [{\citenamefont {Demk\'o}\ \emph {et~al.}(2012)\citenamefont
  {Demk\'o}, \citenamefont {Schober}, \citenamefont {Kocsis}, \citenamefont
  {Bahramy}, \citenamefont {Murakawa}, \citenamefont {Lee}, \citenamefont
  {K\'ezsm\'arki}, \citenamefont {Arita}, \citenamefont {Nagaosa},\ and\
  \citenamefont {Tokura}}]{PhysRevLett.109.167401}%
  \BibitemOpen
  \bibfield  {author} {\bibinfo {author} {\bibfnamefont {L.}~\bibnamefont
  {Demk\'o}}, \bibinfo {author} {\bibfnamefont {G.~A.~H.}\ \bibnamefont
  {Schober}}, \bibinfo {author} {\bibfnamefont {V.}~\bibnamefont {Kocsis}},
  \bibinfo {author} {\bibfnamefont {M.~S.}\ \bibnamefont {Bahramy}}, \bibinfo
  {author} {\bibfnamefont {H.}~\bibnamefont {Murakawa}}, \bibinfo {author}
  {\bibfnamefont {J.~S.}\ \bibnamefont {Lee}}, \bibinfo {author} {\bibfnamefont
  {I.}~\bibnamefont {K\'ezsm\'arki}}, \bibinfo {author} {\bibfnamefont
  {R.}~\bibnamefont {Arita}}, \bibinfo {author} {\bibfnamefont
  {N.}~\bibnamefont {Nagaosa}}, \ and\ \bibinfo {author} {\bibfnamefont
  {Y.}~\bibnamefont {Tokura}},\ }\href {\doibase
  10.1103/PhysRevLett.109.167401} {\bibfield  {journal} {\bibinfo  {journal}
  {Phys. Rev. Lett.}\ }\textbf {\bibinfo {volume} {109}},\ \bibinfo {pages}
  {167401} (\bibinfo {year} {2012})}\BibitemShut {NoStop}%
\bibitem [{\citenamefont {Bahramy}\ \emph {et~al.}(2012)\citenamefont
  {Bahramy}, \citenamefont {Yang}, \citenamefont {Arita},\ and\ \citenamefont
  {Nagaosa}}]{Bahramy2012}%
  \BibitemOpen
  \bibfield  {author} {\bibinfo {author} {\bibfnamefont {M.~S.}\ \bibnamefont
  {Bahramy}}, \bibinfo {author} {\bibfnamefont {B.-J.}\ \bibnamefont {Yang}},
  \bibinfo {author} {\bibfnamefont {R.}~\bibnamefont {Arita}}, \ and\ \bibinfo
  {author} {\bibfnamefont {N.}~\bibnamefont {Nagaosa}},\ }\href
  {https://doi.org/10.1038/ncomms1679} {\bibfield  {journal} {\bibinfo
  {journal} {Nat. Commun.}\ }\textbf {\bibinfo {volume} {3}},\ \bibinfo {pages}
  {679} (\bibinfo {year} {2012})}\BibitemShut {NoStop}%
\bibitem [{\citenamefont {Eremeev}\ \emph {et~al.}(2017)\citenamefont
  {Eremeev}, \citenamefont {Nechaev},\ and\ \citenamefont
  {Chulkov}}]{PhysRevB.96.155309}%
  \BibitemOpen
  \bibfield  {author} {\bibinfo {author} {\bibfnamefont {S.~V.}\ \bibnamefont
  {Eremeev}}, \bibinfo {author} {\bibfnamefont {I.~A.}\ \bibnamefont
  {Nechaev}}, \ and\ \bibinfo {author} {\bibfnamefont {E.~V.}\ \bibnamefont
  {Chulkov}},\ }\href {\doibase 10.1103/PhysRevB.96.155309} {\bibfield
  {journal} {\bibinfo  {journal} {Phys. Rev. B}\ }\textbf {\bibinfo {volume}
  {96}},\ \bibinfo {pages} {155309} (\bibinfo {year} {2017})}\BibitemShut
  {NoStop}%
\bibitem [{\citenamefont {Facio}\ \emph {et~al.}(2018)\citenamefont {Facio},
  \citenamefont {Efremov}, \citenamefont {Koepernik}, \citenamefont {You},
  \citenamefont {Sodemann},\ and\ \citenamefont {van~den
  Brink}}]{PhysRevLett.121.246403}%
  \BibitemOpen
  \bibfield  {author} {\bibinfo {author} {\bibfnamefont {J.~I.}\ \bibnamefont
  {Facio}}, \bibinfo {author} {\bibfnamefont {D.}~\bibnamefont {Efremov}},
  \bibinfo {author} {\bibfnamefont {K.}~\bibnamefont {Koepernik}}, \bibinfo
  {author} {\bibfnamefont {J.-S.}\ \bibnamefont {You}}, \bibinfo {author}
  {\bibfnamefont {I.}~\bibnamefont {Sodemann}}, \ and\ \bibinfo {author}
  {\bibfnamefont {J.}~\bibnamefont {van~den Brink}},\ }\href {\doibase
  10.1103/PhysRevLett.121.246403} {\bibfield  {journal} {\bibinfo  {journal}
  {Phys. Rev. Lett.}\ }\textbf {\bibinfo {volume} {121}},\ \bibinfo {pages}
  {246403} (\bibinfo {year} {2018})}\BibitemShut {NoStop}%
\bibitem [{\citenamefont {Qi}\ \emph {et~al.}(2017)\citenamefont {Qi},
  \citenamefont {Shi}, \citenamefont {Naumov}, \citenamefont {Kumar},
  \citenamefont {Sankar}, \citenamefont {Schnelle}, \citenamefont {Shekhar},
  \citenamefont {Chou}, \citenamefont {Felser}, \citenamefont {Yan},\ and\
  \citenamefont {Medvedev}}]{10.1002/adma.201605965}%
  \BibitemOpen
  \bibfield  {author} {\bibinfo {author} {\bibfnamefont {Y.}~\bibnamefont
  {Qi}}, \bibinfo {author} {\bibfnamefont {W.}~\bibnamefont {Shi}}, \bibinfo
  {author} {\bibfnamefont {P.~G.}\ \bibnamefont {Naumov}}, \bibinfo {author}
  {\bibfnamefont {N.}~\bibnamefont {Kumar}}, \bibinfo {author} {\bibfnamefont
  {R.}~\bibnamefont {Sankar}}, \bibinfo {author} {\bibfnamefont
  {W.}~\bibnamefont {Schnelle}}, \bibinfo {author} {\bibfnamefont
  {C.}~\bibnamefont {Shekhar}}, \bibinfo {author} {\bibfnamefont {F.-C.}\
  \bibnamefont {Chou}}, \bibinfo {author} {\bibfnamefont {C.}~\bibnamefont
  {Felser}}, \bibinfo {author} {\bibfnamefont {B.}~\bibnamefont {Yan}}, \ and\
  \bibinfo {author} {\bibfnamefont {S.~A.}\ \bibnamefont {Medvedev}},\ }\href
  {\doibase 10.1002/adma.201605965} {\bibfield  {journal} {\bibinfo  {journal}
  {Adv. Mater.}\ }\textbf {\bibinfo {volume} {29}},\ \bibinfo {pages} {1605965}
  (\bibinfo {year} {2017})}\BibitemShut {NoStop}%
\bibitem [{\citenamefont {Ma}\ \emph {et~al.}(2014)\citenamefont {Ma},
  \citenamefont {Dai}, \citenamefont {Wei}, \citenamefont {Li},\ and\
  \citenamefont {Huang}}]{ma2014emergence}%
  \BibitemOpen
  \bibfield  {author} {\bibinfo {author} {\bibfnamefont {Y.}~\bibnamefont
  {Ma}}, \bibinfo {author} {\bibfnamefont {Y.}~\bibnamefont {Dai}}, \bibinfo
  {author} {\bibfnamefont {W.}~\bibnamefont {Wei}}, \bibinfo {author}
  {\bibfnamefont {X.}~\bibnamefont {Li}}, \ and\ \bibinfo {author}
  {\bibfnamefont {B.}~\bibnamefont {Huang}},\ }\href@noop {} {\bibfield
  {journal} {\bibinfo  {journal} {Phys. Chem. Chem. Phys.}\ }\textbf {\bibinfo
  {volume} {16}},\ \bibinfo {pages} {17603} (\bibinfo {year}
  {2014})}\BibitemShut {NoStop}%
\bibitem [{\citenamefont {F{\"u}l{\"o}p}\ \emph {et~al.}(2018)\citenamefont
  {F{\"u}l{\"o}p}, \citenamefont {Tajkov}, \citenamefont {Pet{\H{o}}},
  \citenamefont {Kun}, \citenamefont {Koltai}, \citenamefont {Oroszl{\'a}ny},
  \citenamefont {T{\'o}v{\'a}ri}, \citenamefont {Murakawa}, \citenamefont
  {Tokura}, \citenamefont {Bord{\'a}cs} \emph {et~al.}}]{fulop2018exfoliation}%
  \BibitemOpen
  \bibfield  {author} {\bibinfo {author} {\bibfnamefont {B.}~\bibnamefont
  {F{\"u}l{\"o}p}}, \bibinfo {author} {\bibfnamefont {Z.}~\bibnamefont
  {Tajkov}}, \bibinfo {author} {\bibfnamefont {J.}~\bibnamefont {Pet{\H{o}}}},
  \bibinfo {author} {\bibfnamefont {P.}~\bibnamefont {Kun}}, \bibinfo {author}
  {\bibfnamefont {J.}~\bibnamefont {Koltai}}, \bibinfo {author} {\bibfnamefont
  {L.}~\bibnamefont {Oroszl{\'a}ny}}, \bibinfo {author} {\bibfnamefont
  {E.}~\bibnamefont {T{\'o}v{\'a}ri}}, \bibinfo {author} {\bibfnamefont
  {H.}~\bibnamefont {Murakawa}}, \bibinfo {author} {\bibfnamefont
  {Y.}~\bibnamefont {Tokura}}, \bibinfo {author} {\bibfnamefont
  {S.}~\bibnamefont {Bord{\'a}cs}},  \emph {et~al.},\ }\href@noop {} {\bibfield
   {journal} {\bibinfo  {journal} {2D Mater.}\ }\textbf {\bibinfo {volume}
  {5}},\ \bibinfo {pages} {031013} (\bibinfo {year} {2018})}\BibitemShut
  {NoStop}%
\bibitem [{\citenamefont {Bl\"ochl}(1994)}]{PhysRevB.50.17953}%
  \BibitemOpen
  \bibfield  {author} {\bibinfo {author} {\bibfnamefont {P.~E.}\ \bibnamefont
  {Bl\"ochl}},\ }\href {\doibase 10.1103/PhysRevB.50.17953} {\bibfield
  {journal} {\bibinfo  {journal} {Phys. Rev. B}\ }\textbf {\bibinfo {volume}
  {50}},\ \bibinfo {pages} {17953} (\bibinfo {year} {1994})}\BibitemShut
  {NoStop}%
\bibitem [{\citenamefont {Kresse}\ and\ \citenamefont
  {Hafner}(1993)}]{PhysRevB.47.558}%
  \BibitemOpen
  \bibfield  {author} {\bibinfo {author} {\bibfnamefont {G.}~\bibnamefont
  {Kresse}}\ and\ \bibinfo {author} {\bibfnamefont {J.}~\bibnamefont
  {Hafner}},\ }\href {\doibase 10.1103/PhysRevB.47.558} {\bibfield  {journal}
  {\bibinfo  {journal} {Phys. Rev. B}\ }\textbf {\bibinfo {volume} {47}},\
  \bibinfo {pages} {558} (\bibinfo {year} {1993})}\BibitemShut {NoStop}%
\bibitem [{\citenamefont {Perdew}\ \emph {et~al.}(1996)\citenamefont {Perdew},
  \citenamefont {Burke},\ and\ \citenamefont
  {Ernzerhof}}]{PhysRevLett.77.3865}%
  \BibitemOpen
  \bibfield  {author} {\bibinfo {author} {\bibfnamefont {J.~P.}\ \bibnamefont
  {Perdew}}, \bibinfo {author} {\bibfnamefont {K.}~\bibnamefont {Burke}}, \
  and\ \bibinfo {author} {\bibfnamefont {M.}~\bibnamefont {Ernzerhof}},\ }\href
  {\doibase 10.1103/PhysRevLett.77.3865} {\bibfield  {journal} {\bibinfo
  {journal} {Phys. Rev. Lett.}\ }\textbf {\bibinfo {volume} {77}},\ \bibinfo
  {pages} {3865} (\bibinfo {year} {1996})}\BibitemShut {NoStop}%
\bibitem [{\citenamefont {Monkhorst}\ and\ \citenamefont
  {Pack}(1976)}]{PhysRevB.13.5188}%
  \BibitemOpen
  \bibfield  {author} {\bibinfo {author} {\bibfnamefont {H.~J.}\ \bibnamefont
  {Monkhorst}}\ and\ \bibinfo {author} {\bibfnamefont {J.~D.}\ \bibnamefont
  {Pack}},\ }\href {\doibase 10.1103/PhysRevB.13.5188} {\bibfield  {journal}
  {\bibinfo  {journal} {Phys. Rev. B}\ }\textbf {\bibinfo {volume} {13}},\
  \bibinfo {pages} {5188} (\bibinfo {year} {1976})}\BibitemShut {NoStop}%
\bibitem [{\citenamefont {Vajna}\ \emph {et~al.}(2012)\citenamefont {Vajna},
  \citenamefont {Simon}, \citenamefont {Szilva}, \citenamefont {Palotas},
  \citenamefont {Ujfalussy},\ and\ \citenamefont
  {Szunyogh}}]{PhysRevB.85.075404}%
  \BibitemOpen
  \bibfield  {author} {\bibinfo {author} {\bibfnamefont {S.}~\bibnamefont
  {Vajna}}, \bibinfo {author} {\bibfnamefont {E.}~\bibnamefont {Simon}},
  \bibinfo {author} {\bibfnamefont {A.}~\bibnamefont {Szilva}}, \bibinfo
  {author} {\bibfnamefont {K.}~\bibnamefont {Palotas}}, \bibinfo {author}
  {\bibfnamefont {B.}~\bibnamefont {Ujfalussy}}, \ and\ \bibinfo {author}
  {\bibfnamefont {L.}~\bibnamefont {Szunyogh}},\ }\href {\doibase
  10.1103/PhysRevB.85.075404} {\bibfield  {journal} {\bibinfo  {journal} {Phys.
  Rev. B}\ }\textbf {\bibinfo {volume} {85}},\ \bibinfo {pages} {075404}
  (\bibinfo {year} {2012})}\BibitemShut {NoStop}%
\bibitem [{\citenamefont {Fu}(2009)}]{PhysRevLett.103.266801}%
  \BibitemOpen
  \bibfield  {author} {\bibinfo {author} {\bibfnamefont {L.}~\bibnamefont
  {Fu}},\ }\href {\doibase 10.1103/PhysRevLett.103.266801} {\bibfield
  {journal} {\bibinfo  {journal} {Phys. Rev. Lett.}\ }\textbf {\bibinfo
  {volume} {103}},\ \bibinfo {pages} {266801} (\bibinfo {year}
  {2009})}\BibitemShut {NoStop}%
\bibitem [{\citenamefont {Yu}\ \emph {et~al.}(2014)\citenamefont {Yu},
  \citenamefont {Wu}, \citenamefont {Liu}, \citenamefont {Xu},\ and\
  \citenamefont {Yao}}]{PhysRevLett.113.156603}%
  \BibitemOpen
  \bibfield  {author} {\bibinfo {author} {\bibfnamefont {H.}~\bibnamefont
  {Yu}}, \bibinfo {author} {\bibfnamefont {Y.}~\bibnamefont {Wu}}, \bibinfo
  {author} {\bibfnamefont {G.-B.}\ \bibnamefont {Liu}}, \bibinfo {author}
  {\bibfnamefont {X.}~\bibnamefont {Xu}}, \ and\ \bibinfo {author}
  {\bibfnamefont {W.}~\bibnamefont {Yao}},\ }\href {\doibase
  10.1103/PhysRevLett.113.156603} {\bibfield  {journal} {\bibinfo  {journal}
  {Phys. Rev. Lett.}\ }\textbf {\bibinfo {volume} {113}},\ \bibinfo {pages}
  {156603} (\bibinfo {year} {2014})}\BibitemShut {NoStop}%
\bibitem [{\citenamefont {Hamamoto}\ \emph {et~al.}(2017)\citenamefont
  {Hamamoto}, \citenamefont {Ezawa}, \citenamefont {Kim}, \citenamefont
  {Morimoto},\ and\ \citenamefont {Nagaosa}}]{PhysRevB.95.224430}%
  \BibitemOpen
  \bibfield  {author} {\bibinfo {author} {\bibfnamefont {K.}~\bibnamefont
  {Hamamoto}}, \bibinfo {author} {\bibfnamefont {M.}~\bibnamefont {Ezawa}},
  \bibinfo {author} {\bibfnamefont {K.~W.}\ \bibnamefont {Kim}}, \bibinfo
  {author} {\bibfnamefont {T.}~\bibnamefont {Morimoto}}, \ and\ \bibinfo
  {author} {\bibfnamefont {N.}~\bibnamefont {Nagaosa}},\ }\href {\doibase
  10.1103/PhysRevB.95.224430} {\bibfield  {journal} {\bibinfo  {journal} {Phys.
  Rev. B}\ }\textbf {\bibinfo {volume} {95}},\ \bibinfo {pages} {224430}
  (\bibinfo {year} {2017})}\BibitemShut {NoStop}%
\bibitem [{\citenamefont {Sodemann}\ and\ \citenamefont
  {Fu}(2015)}]{PhysRevLett.115.216806}%
  \BibitemOpen
  \bibfield  {author} {\bibinfo {author} {\bibfnamefont {I.}~\bibnamefont
  {Sodemann}}\ and\ \bibinfo {author} {\bibfnamefont {L.}~\bibnamefont {Fu}},\
  }\href {\doibase 10.1103/PhysRevLett.115.216806} {\bibfield  {journal}
  {\bibinfo  {journal} {Phys. Rev. Lett.}\ }\textbf {\bibinfo {volume} {115}},\
  \bibinfo {pages} {216806} (\bibinfo {year} {2015})}\BibitemShut {NoStop}%
\bibitem [{\citenamefont {Morimoto}\ \emph {et~al.}(2016)\citenamefont
  {Morimoto}, \citenamefont {Zhong}, \citenamefont {Orenstein},\ and\
  \citenamefont {Moore}}]{PhysRevB.94.245121}%
  \BibitemOpen
  \bibfield  {author} {\bibinfo {author} {\bibfnamefont {T.}~\bibnamefont
  {Morimoto}}, \bibinfo {author} {\bibfnamefont {S.}~\bibnamefont {Zhong}},
  \bibinfo {author} {\bibfnamefont {J.}~\bibnamefont {Orenstein}}, \ and\
  \bibinfo {author} {\bibfnamefont {J.~E.}\ \bibnamefont {Moore}},\ }\href
  {\doibase 10.1103/PhysRevB.94.245121} {\bibfield  {journal} {\bibinfo
  {journal} {Phys. Rev. B}\ }\textbf {\bibinfo {volume} {94}},\ \bibinfo
  {pages} {245121} (\bibinfo {year} {2016})}\BibitemShut {NoStop}%
\bibitem [{\citenamefont {Rashba}(2003)}]{PhysRevB.68.241315}%
  \BibitemOpen
  \bibfield  {author} {\bibinfo {author} {\bibfnamefont {E.~I.}\ \bibnamefont
  {Rashba}},\ }\href {\doibase 10.1103/PhysRevB.68.241315} {\bibfield
  {journal} {\bibinfo  {journal} {Phys. Rev. B}\ }\textbf {\bibinfo {volume}
  {68}},\ \bibinfo {pages} {241315} (\bibinfo {year} {2003})}\BibitemShut
  {NoStop}%
\bibitem [{\citenamefont {Sinova}\ \emph {et~al.}(2004)\citenamefont {Sinova},
  \citenamefont {Culcer}, \citenamefont {Niu}, \citenamefont {Sinitsyn},
  \citenamefont {Jungwirth},\ and\ \citenamefont
  {MacDonald}}]{PhysRevLett.92.126603}%
  \BibitemOpen
  \bibfield  {author} {\bibinfo {author} {\bibfnamefont {J.}~\bibnamefont
  {Sinova}}, \bibinfo {author} {\bibfnamefont {D.}~\bibnamefont {Culcer}},
  \bibinfo {author} {\bibfnamefont {Q.}~\bibnamefont {Niu}}, \bibinfo {author}
  {\bibfnamefont {N.~A.}\ \bibnamefont {Sinitsyn}}, \bibinfo {author}
  {\bibfnamefont {T.}~\bibnamefont {Jungwirth}}, \ and\ \bibinfo {author}
  {\bibfnamefont {A.~H.}\ \bibnamefont {MacDonald}},\ }\href {\doibase
  10.1103/PhysRevLett.92.126603} {\bibfield  {journal} {\bibinfo  {journal}
  {Phys. Rev. Lett.}\ }\textbf {\bibinfo {volume} {92}},\ \bibinfo {pages}
  {126603} (\bibinfo {year} {2004})}\BibitemShut {NoStop}%
\bibitem [{\citenamefont {Drouhin}\ \emph {et~al.}(2011)\citenamefont
  {Drouhin}, \citenamefont {Fishman},\ and\ \citenamefont
  {Wegrowe}}]{PhysRevB.83.113307}%
  \BibitemOpen
  \bibfield  {author} {\bibinfo {author} {\bibfnamefont {H.-J.}\ \bibnamefont
  {Drouhin}}, \bibinfo {author} {\bibfnamefont {G.}~\bibnamefont {Fishman}}, \
  and\ \bibinfo {author} {\bibfnamefont {J.-E.}\ \bibnamefont {Wegrowe}},\
  }\href {\doibase 10.1103/PhysRevB.83.113307} {\bibfield  {journal} {\bibinfo
  {journal} {Phys. Rev. B}\ }\textbf {\bibinfo {volume} {83}},\ \bibinfo
  {pages} {113307} (\bibinfo {year} {2011})}\BibitemShut {NoStop}%
\bibitem [{\citenamefont {Tokatly}(2008)}]{PhysRevLett.101.106601}%
  \BibitemOpen
  \bibfield  {author} {\bibinfo {author} {\bibfnamefont {I.~V.}\ \bibnamefont
  {Tokatly}},\ }\href {\doibase 10.1103/PhysRevLett.101.106601} {\bibfield
  {journal} {\bibinfo  {journal} {Phys. Rev. Lett.}\ }\textbf {\bibinfo
  {volume} {101}},\ \bibinfo {pages} {106601} (\bibinfo {year}
  {2008})}\BibitemShut {NoStop}%
\bibitem [{\citenamefont {Sonin}(2007)}]{PhysRevB.76.033306}%
  \BibitemOpen
  \bibfield  {author} {\bibinfo {author} {\bibfnamefont {E.~B.}\ \bibnamefont
  {Sonin}},\ }\href {\doibase 10.1103/PhysRevB.76.033306} {\bibfield  {journal}
  {\bibinfo  {journal} {Phys. Rev. B}\ }\textbf {\bibinfo {volume} {76}},\
  \bibinfo {pages} {033306} (\bibinfo {year} {2007})}\BibitemShut {NoStop}%
\bibitem [{\citenamefont {Sun}\ \emph {et~al.}(2008)\citenamefont {Sun},
  \citenamefont {Xie},\ and\ \citenamefont {Wang}}]{PhysRevB.77.035327}%
  \BibitemOpen
  \bibfield  {author} {\bibinfo {author} {\bibfnamefont {Q.-F.}\ \bibnamefont
  {Sun}}, \bibinfo {author} {\bibfnamefont {X.~C.}\ \bibnamefont {Xie}}, \ and\
  \bibinfo {author} {\bibfnamefont {J.}~\bibnamefont {Wang}},\ }\href {\doibase
  10.1103/PhysRevB.77.035327} {\bibfield  {journal} {\bibinfo  {journal} {Phys.
  Rev. B}\ }\textbf {\bibinfo {volume} {77}},\ \bibinfo {pages} {035327}
  (\bibinfo {year} {2008})}\BibitemShut {NoStop}%
\bibitem [{\citenamefont {Martin}\ \emph {et~al.}(2013)\citenamefont {Martin},
  \citenamefont {Mun}, \citenamefont {Berger}, \citenamefont {Zapf},\ and\
  \citenamefont {Tanner}}]{PhysRevB.87.041104}%
  \BibitemOpen
  \bibfield  {author} {\bibinfo {author} {\bibfnamefont {C.}~\bibnamefont
  {Martin}}, \bibinfo {author} {\bibfnamefont {E.~D.}\ \bibnamefont {Mun}},
  \bibinfo {author} {\bibfnamefont {H.}~\bibnamefont {Berger}}, \bibinfo
  {author} {\bibfnamefont {V.~S.}\ \bibnamefont {Zapf}}, \ and\ \bibinfo
  {author} {\bibfnamefont {D.~B.}\ \bibnamefont {Tanner}},\ }\href {\doibase
  10.1103/PhysRevB.87.041104} {\bibfield  {journal} {\bibinfo  {journal} {Phys.
  Rev. B}\ }\textbf {\bibinfo {volume} {87}},\ \bibinfo {pages} {041104}
  (\bibinfo {year} {2013})}\BibitemShut {NoStop}%
\bibitem [{\citenamefont {Mak}\ \emph {et~al.}(2013)\citenamefont {Mak},
  \citenamefont {He}, \citenamefont {Lee}, \citenamefont {Lee}, \citenamefont
  {Hone}, \citenamefont {Heinz},\ and\ \citenamefont {Shan}}]{mak2013tightly}%
  \BibitemOpen
  \bibfield  {author} {\bibinfo {author} {\bibfnamefont {K.~F.}\ \bibnamefont
  {Mak}}, \bibinfo {author} {\bibfnamefont {K.}~\bibnamefont {He}}, \bibinfo
  {author} {\bibfnamefont {C.}~\bibnamefont {Lee}}, \bibinfo {author}
  {\bibfnamefont {G.~H.}\ \bibnamefont {Lee}}, \bibinfo {author} {\bibfnamefont
  {J.}~\bibnamefont {Hone}}, \bibinfo {author} {\bibfnamefont {T.~F.}\
  \bibnamefont {Heinz}}, \ and\ \bibinfo {author} {\bibfnamefont
  {J.}~\bibnamefont {Shan}},\ }\href@noop {} {\bibfield  {journal} {\bibinfo
  {journal} {Nat. Mater.}\ }\textbf {\bibinfo {volume} {12}},\ \bibinfo {pages}
  {207} (\bibinfo {year} {2013})}\BibitemShut {NoStop}%
\bibitem [{\citenamefont {Zhang}\ \emph {et~al.}(2014)\citenamefont {Zhang},
  \citenamefont {Oka}, \citenamefont {Suzuki}, \citenamefont {Ye},\ and\
  \citenamefont {Iwasa}}]{zhang2014electrically}%
  \BibitemOpen
  \bibfield  {author} {\bibinfo {author} {\bibfnamefont {Y.}~\bibnamefont
  {Zhang}}, \bibinfo {author} {\bibfnamefont {T.}~\bibnamefont {Oka}}, \bibinfo
  {author} {\bibfnamefont {R.}~\bibnamefont {Suzuki}}, \bibinfo {author}
  {\bibfnamefont {J.}~\bibnamefont {Ye}}, \ and\ \bibinfo {author}
  {\bibfnamefont {Y.}~\bibnamefont {Iwasa}},\ }\href@noop {} {\bibfield
  {journal} {\bibinfo  {journal} {Science}\ }\textbf {\bibinfo {volume}
  {344}},\ \bibinfo {pages} {725} (\bibinfo {year} {2014})}\BibitemShut
  {NoStop}%
\bibitem [{\citenamefont {Efetov}\ and\ \citenamefont
  {Kim}(2010)}]{PhysRevLett.105.256805}%
  \BibitemOpen
  \bibfield  {author} {\bibinfo {author} {\bibfnamefont {D.~K.}\ \bibnamefont
  {Efetov}}\ and\ \bibinfo {author} {\bibfnamefont {P.}~\bibnamefont {Kim}},\
  }\href {\doibase 10.1103/PhysRevLett.105.256805} {\bibfield  {journal}
  {\bibinfo  {journal} {Phys. Rev. Lett.}\ }\textbf {\bibinfo {volume} {105}},\
  \bibinfo {pages} {256805} (\bibinfo {year} {2010})}\BibitemShut {NoStop}%
\bibitem [{\citenamefont {Ye}\ \emph {et~al.}(2011)\citenamefont {Ye},
  \citenamefont {Craciun}, \citenamefont {Koshino}, \citenamefont {Russo},
  \citenamefont {Inoue}, \citenamefont {Yuan}, \citenamefont {Shimotani},
  \citenamefont {Morpurgo},\ and\ \citenamefont {Iwasa}}]{Ye13002}%
  \BibitemOpen
  \bibfield  {author} {\bibinfo {author} {\bibfnamefont {J.}~\bibnamefont
  {Ye}}, \bibinfo {author} {\bibfnamefont {M.~F.}\ \bibnamefont {Craciun}},
  \bibinfo {author} {\bibfnamefont {M.}~\bibnamefont {Koshino}}, \bibinfo
  {author} {\bibfnamefont {S.}~\bibnamefont {Russo}}, \bibinfo {author}
  {\bibfnamefont {S.}~\bibnamefont {Inoue}}, \bibinfo {author} {\bibfnamefont
  {H.}~\bibnamefont {Yuan}}, \bibinfo {author} {\bibfnamefont {H.}~\bibnamefont
  {Shimotani}}, \bibinfo {author} {\bibfnamefont {A.~F.}\ \bibnamefont
  {Morpurgo}}, \ and\ \bibinfo {author} {\bibfnamefont {Y.}~\bibnamefont
  {Iwasa}},\ }\href {\doibase 10.1073/pnas.1018388108} {\bibfield  {journal}
  {\bibinfo  {journal} {Proc. Natl. Acad. Sci.}\ }\textbf {\bibinfo {volume}
  {108}},\ \bibinfo {pages} {13002} (\bibinfo {year} {2011})}\BibitemShut
  {NoStop}%
\bibitem [{\citenamefont {Zhou}\ \emph {et~al.}(2018)\citenamefont {Zhou},
  \citenamefont {Lin}, \citenamefont {Huang}, \citenamefont {Zhou},
  \citenamefont {Chen}, \citenamefont {Xia}, \citenamefont {Wang},
  \citenamefont {Xie}, \citenamefont {Yu}, \citenamefont {Lei}, \citenamefont
  {Wu}, \citenamefont {Liu}, \citenamefont {Fu}, \citenamefont {Zeng},
  \citenamefont {Hsu}, \citenamefont {Yang}, \citenamefont {Lu}, \citenamefont
  {Yu}, \citenamefont {Shen}, \citenamefont {Lin}, \citenamefont {Yakobson},
  \citenamefont {Liu}, \citenamefont {Suenaga}, \citenamefont {Liu},\ and\
  \citenamefont {Liu}}]{Zhou2018}%
  \BibitemOpen
  \bibfield  {author} {\bibinfo {author} {\bibfnamefont {J.}~\bibnamefont
  {Zhou}}, \bibinfo {author} {\bibfnamefont {J.}~\bibnamefont {Lin}}, \bibinfo
  {author} {\bibfnamefont {X.}~\bibnamefont {Huang}}, \bibinfo {author}
  {\bibfnamefont {Y.}~\bibnamefont {Zhou}}, \bibinfo {author} {\bibfnamefont
  {Y.}~\bibnamefont {Chen}}, \bibinfo {author} {\bibfnamefont {J.}~\bibnamefont
  {Xia}}, \bibinfo {author} {\bibfnamefont {H.}~\bibnamefont {Wang}}, \bibinfo
  {author} {\bibfnamefont {Y.}~\bibnamefont {Xie}}, \bibinfo {author}
  {\bibfnamefont {H.}~\bibnamefont {Yu}}, \bibinfo {author} {\bibfnamefont
  {J.}~\bibnamefont {Lei}}, \bibinfo {author} {\bibfnamefont {D.}~\bibnamefont
  {Wu}}, \bibinfo {author} {\bibfnamefont {F.}~\bibnamefont {Liu}}, \bibinfo
  {author} {\bibfnamefont {Q.}~\bibnamefont {Fu}}, \bibinfo {author}
  {\bibfnamefont {Q.}~\bibnamefont {Zeng}}, \bibinfo {author} {\bibfnamefont
  {C.-H.}\ \bibnamefont {Hsu}}, \bibinfo {author} {\bibfnamefont
  {C.}~\bibnamefont {Yang}}, \bibinfo {author} {\bibfnamefont {L.}~\bibnamefont
  {Lu}}, \bibinfo {author} {\bibfnamefont {T.}~\bibnamefont {Yu}}, \bibinfo
  {author} {\bibfnamefont {Z.}~\bibnamefont {Shen}}, \bibinfo {author}
  {\bibfnamefont {H.}~\bibnamefont {Lin}}, \bibinfo {author} {\bibfnamefont
  {B.~I.}\ \bibnamefont {Yakobson}}, \bibinfo {author} {\bibfnamefont
  {Q.}~\bibnamefont {Liu}}, \bibinfo {author} {\bibfnamefont {K.}~\bibnamefont
  {Suenaga}}, \bibinfo {author} {\bibfnamefont {G.}~\bibnamefont {Liu}}, \ and\
  \bibinfo {author} {\bibfnamefont {Z.}~\bibnamefont {Liu}},\ }\href {\doibase
  10.1038/s41586-018-0008-3} {\bibfield  {journal} {\bibinfo  {journal}
  {Nature}\ }\textbf {\bibinfo {volume} {556}},\ \bibinfo {pages} {355}
  (\bibinfo {year} {2018})}\BibitemShut {NoStop}%
\end{thebibliography}

%

\end{document}